\documentclass[useAMS,usenatbib]{mn2e}
\usepackage{amsmath,amsfonts,amssymb,theorem,graphics }

\begin{document}

\title[Time dependent non-LTE calculations of ionisation in the early universe]
{Time dependent non-LTE calculations of ionisation in the early universe}  
\author[R. Wehrse, D. T. Wickramasinghe, and R. Dav\'e]
{R. Wehrse$^{1}$\thanks{E-mail: wehrse@ita.uni-heidelberg.de};
D.T. Wickramasinghe$^{2}$;
R. Dav\'e$^{3}$
\\
$^{1}$Zentrum f. Astronomie Heidelberg, Institut f. Theoret. Astrophysik, A.-Ueberle-Str.2, D 69120 Heidelberg, 
Germany\\
$^{2}$The National University, Canberra ACT 2617, Australia\\
$^{3}$Steward Observatory, University of Arizona, Tucson, USA\\}

\date{Received 2005 July}
\pagerange{\pageref{firstpage}--\pageref{lastpage}} \pubyear{2002}
\maketitle
\label{firstpage}

\begin {abstract}
We present a new implicit numerical algorithm for the calculation of the time 
dependent non-Local Thermodynamic Equilibrium of a gas in an external 
radiation field that is accurate, fast and unconditionally   
stable for all spatial and temporal increments. The method is presented 
as a backward difference scheme in 1-D but can be readily generalised to 3-D.
We apply the method for calculating the evolution of ionisation domains in a
hydrogen plasma with plane-parallel Gaussian density enhancements illuminated
by sources of UV radiation. We calculate the speed of propagation of 
ionising fronts through different ambient densities and the interaction of 
such ionising fronts with density enhancements. We show that for a
typical UV source that may be present in the early universe, the 
introduction of a density 
enhancement of a factor $\sim 10$ above an ambient density  
$10^{-4}$ cm$^{-3}$ could delay the outward propagation of an ionisation front 
by millions of years. Our calculations show that within the lifetime of a 
single source ($\sim$ a few million years), and for ambient intergalactic 
densities 
appropriate to redshifts $z\sim 6 - 20$, degrees of ionisation of
$\sim 10^{-3} -10^{-5}$ can be achieved within its zone of influence. 
We also present calculations
which demonstrate that once started, ionisation will proceed
very efficiently as multiple sources are subsequently introduced,
even if the time between the appearence of such sources may be much
longer than their lifetimes. 
 
\end{abstract}

\section{introduction}
In local thermodynamic equilibrium it is explicitly assumed that  the local 
kinetic temperature of the gas 
determines the level populations of the atoms or molecules via the Boltzmann 
distribution. This situation 
pertains when collisions dominate or when the radiation field is a black body 
corresponding to the local 
kinetic temperature of the gas. In contrast, in non-LTE, the level populations 
are determined by the rate equations,
and depend both on the local kinetic temperature and on the local radiation 
field which may not be black body like
(see Oxenius (1986) for  precise definitions).  In most problems in 
astrophysics, non-LTE is associated with 
steady state level populations.

There are situations where time dependent non-LTE calculations are required. 
This could occur for instance  
when the hydrodynamical time scales are shorter or comparable to the radiative 
time scales (e.g. in the formation 
of  shock fronts or in stellar explosions).  Another situation of current 
interest where time dependent  effects 
are important occurs in the description of the re-ionisation of the universe  
(e.g. Peacock (2000), Liddle and Lyth 
(2000)).  Here, the densities are sufficiently low and the recombination times 
correspondingly large that a global 
equilibrium between photoionisation and recombination may not be achieved during 
the lifetime of the source of the
ionising photons.  The modelling of this phenomenon in an inhomogeneous medium 
with a complex 3D geometry is a
formidable computational task.

The past few years have seen the development of many numerical methods aimed at 
addressing
the problem of re-ionisation in a 3-D framework. Rather than attempting to solve 
this problem in its
full complexity, many of these methods make practical approximations to the 
solution of the 3-D transfer equation.  
We note in particular adaptive ray tracing methods (Abel, Norman \& Madau 1999, 
Abel \& Wandelt  2002, Sokasian et al. 2003),
the fast Fourier transform method (Cen 2002) and the moment equation method 
using Eddington tensors with an optically
thin approximation (Gnedin \&Abel 2001). In all of these methods, a key element 
is the coupling between the
chemical (rate) equations and the radiative transfer equations which  determines 
how well the ionisation front 
is resolved.  In general no explicit allowance is made for the time rate of 
change of the 
specific intensity (the $\frac{1}{c}\frac{\partial I}{\partial t}$ term)
in the transfer equation so that the methods yield faster than light propagation 
of ionisation fronts close to
the source. Although these methods cannot be rigorously justified in all their 
assumptions, they are physically
motivated,  and provide a description of the process of re-ionisation at some 
level.   Another approach has been
to use Monte Carlo methods (e.g. Masselli, Ferrara \& Ciadi 2003,  Masselli et 
al. 2004). The major limitation
here is that they require the adoption of an explicit time stepping method which 
is unstable for large time steps. This, combined with the large number 
of trials that are needed, appears to limit the method by the computational
effort that is required.
   
In this paper we present a new and robust numerical method for solving the time 
dependent rate equations
and the time dependent radiative transfer equation concurrently using a backward 
differencing scheme.
In section 2 we discuss the nature of the problem and present our numerical 
scheme formulated in 1-spatial dimension.   
Illustrative results describing the time evolution of ionisation domains that 
propagate through  homogeneous and inhomogeneous 
plane parallel distributions of initially neutral hydrogen gas  are presented in 
section 3.  The effects of including multiple sources are discussed in 
section 4. The models  in these 
two sections can serve as a useful bench mark 
for comparisons of results from other codes and for understanding the process of 
re-ionisation in the early universe.  
Our main results are summarised in section 5. 
 
\section{The basic equations and numerical method}
 
We consider only the simplest case of the  re-ionisation of hydrogen although
the method can be easily extended to include helium. The basic 
equations are then the rate equations
which describe the populations of  neutral hydrogen (HI) and ionised hydrogen 
(HII), and the radiative transfer
equation which describes 
the evolution of the mean radiation intensity due to the absorption and the 
emission of radiation. 
\subsection{Formulation for a two level atom}

In the simplest case that we consider, the  model H atoms have one bound 
and one continuum state. The two processes 
of relevance are then ionization from the lower (ground) state and recombination 
from the upper (continuum) state.
 
The rate equation for the number density of  neutral hydrogen atoms ${\it N_0}$  
can be written as
\begin{equation}
\label{rateeq}
\frac{\partial{\it N}_0( x,t)}{\partial t} = -  \sigma J(x,t)  {\it 
N}_0({x},t)+\alpha ({\it N_{\rm tot}}- {\it N}_0({ x},t))^2
\end{equation}
where $N_{\rm tot}=N_0+N_1$ and $N_1$ is the number density of ionised hydrogen 
(HII). 
Here $\sigma$ is the photoionisation cross section,  $\alpha$ is the 
recombination coefficient, and 
\begin{equation} 
J(x,t)=\frac{1}{4\pi}\int_{4\pi}I(x,t,{\bf n})d\omega
\end{equation}
is the mean intensity of the radiation field where $d\omega$ is the element of 
solid angle. 
Note that all occupation numbers are functions of position $x$ and time $t$.  
Since the radiative rates depend 
on the local mean intensity, it is necessary also to solve concurrently the 
radiative transfer equation which we give here for a ray in the direction ${\bf n}=(1,0,0)$ 
\begin{eqnarray}
\label{transfeq}
&& \frac{1}{c}\frac{\partial I(x,t)}{\partial t} + \frac{\partial 
I(x,t)}{\partial x} =-\sigma{\it N_0}(x,t)\left( I(x,t)
-S(x,t)\right). 
\end{eqnarray}
where $S(x,t)$ is an appropriate source function. 
Equations \ref{rateeq} and \ref{transfeq} have to be solved as an initial value 
problem with 
${\it N}_0(x,0)={\it N}_{tot}(x),{\it N}_1(x,0)=0$ with the inflow boundary 
condition 
$I(0,t) = I_{irr}$ on the radiation field where $I_{irr}$ is specified.  That 
is, we assume that the  
medium is completely neutral at the time $t=0$ when the source of ionising 
photons first 
turns on.

\begin{table*}
 \centering
 \begin{minipage}{140mm}
\caption{Basic model data}
\begin{tabular}{lrrrrrrrr}
  \hline
Model   & ${\it N}_{\rm amb}$ & ${\it N}_{\rm enh}$ & $x_s$ & ${\it l}_{\rm w}$ & 
$I_{\rm {irr}}$ & $\Sigma_{\rm start}^0$ &  $v_1$,$v_2$,$v_3$ & $\Delta t$ 
\\
        & cm$^{-3}$       &   cm$^{-3}$ & 10$^{21}$ cm & 10$^{21}$ cm & 10$^5$ \rm{erg}~ \rm {cm$^{-2}$} \rm s$^{-1}$        &   
10$^{18}$ cm$^{-2}$     & 10$^{9}$ cm s${-1}$     &  10$^{13}$ s 
\\
\hline
a & 10$^{-5}$ & $0$ & - & - & 1.0& $0.00064$&     $1.3$  &  \\
b & 10$^{-4}$ & $0$ & - & - & 1.0& $0.0064$ &     $0.96 $&  \\
c & 10$^{-3}$ & $0$ & - & - & 1.0& $0.064$  &     $0.20 $&  \\
d & 10$^{-3}$ & $0$ & - & - & 3.0& $0.064$  &     $0.30 $&  \\
  &           &     &   &   &          &          &    &    \\
b$_1$ & 8.61~10$^{-5}$ &  9.47~10$^{-4}$ & 31 & 0.5 &1.0&  $0.0064$ &   $1.1$;$1.2$;$0$ & 8      
\\
b$_2$ & 7.83~10$^{-5}$ &  8.61~10$^{-4}$ & 31 & 2.0 &1.0&  $0.0064$ &   $2.0$;$1.7$;$0.07$ & 26  
\\
b$_3$ & 4.74~10$^{-5}$ &  4.93~10$^{-4}$ & 31 & 4.0 &1.0&  $0.0064$ &  $1.8$,$2.5$;$0.13$ & 31  \\
  &           &     &  &          &    & &          \\
b$_4$ & 4.94~10$^{-5}$ &  5.12~10$^{-4}$ & 0& 4.0  &1.0& $0.0064$ &     $0.55$;$0.67$  \\
  &           &     &   &          &   &  &          \\
s$_1$(two sources) & 10$^{-4}$ & 10$^{-3}$ & - & 4.0 &1.0& $0.0064$  &     $ -$&  \\
\\  
\hline
\end{tabular}
\end{minipage}
\end{table*}

\begin{figure}
\center
\scalebox{0.6}{\includegraphics{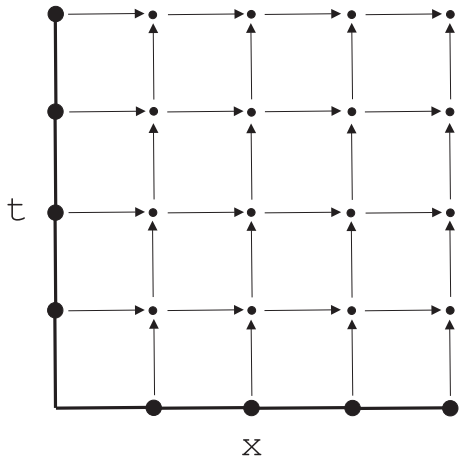}}
\caption{\label{grid_march} Sequence of steps to be used to cover the $x,t$ 
plane. The starting
position is at the lower left corner. The thick lines and large dots indicate 
the where the
initial condition are to be specified.
}
\end{figure}

\begin{figure}
\center
\scalebox{1.1}{\includegraphics{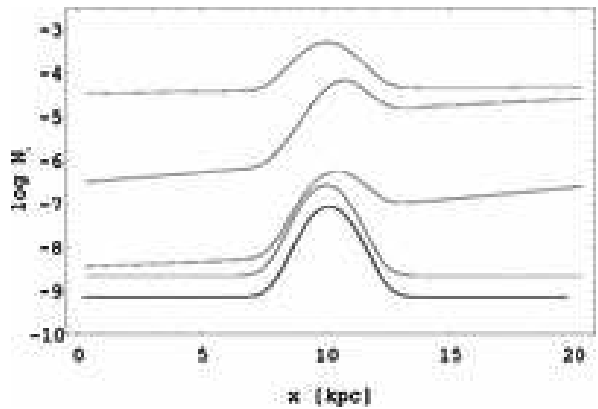}}
\caption{\label{istcomp} 
The convergence to the stationary state. The curves show from top to bottom
the evolution of the Model b3 (see Table 1) density distribution after $0, 2, 4, 8$ Ma
(assuming that the illuminating source lives for this period) and the corresponding
steady state distribution calculated by a completely 
different algorithm (see text). For clarity the latter is shifted by $\Delta \log N_0 =-0.5$,
otherwise the two lowest curves would overlap.
} 
\end{figure}

\begin{figure}
\scalebox{0.90}{\includegraphics{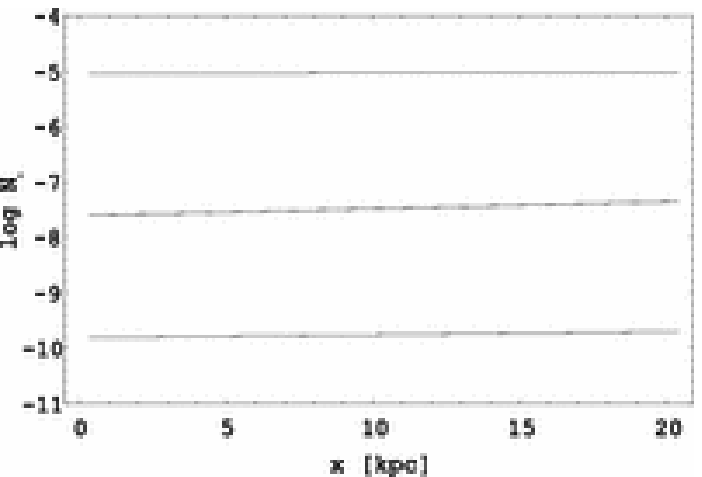}}
\vspace*{0.5cm}

\scalebox{0.9}{\includegraphics{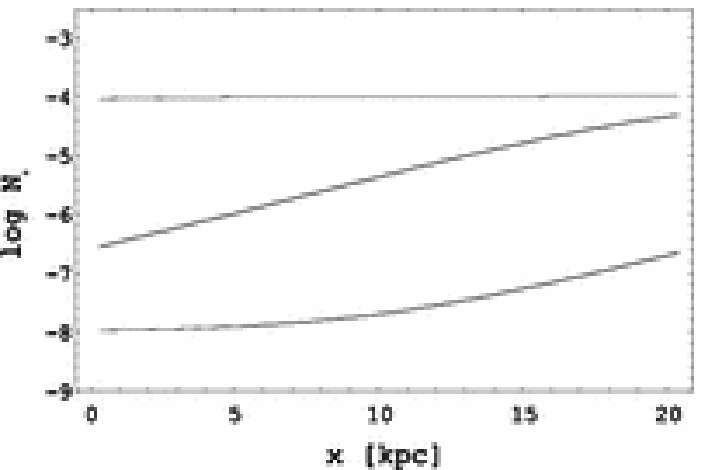}}
\vspace*{0.5cm}

\scalebox{0.9}{\includegraphics{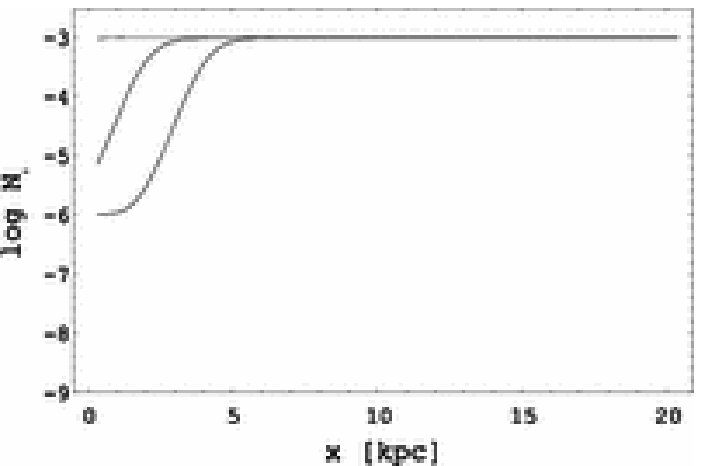}}
\vspace*{0.5cm}

\scalebox{0.90}{\includegraphics{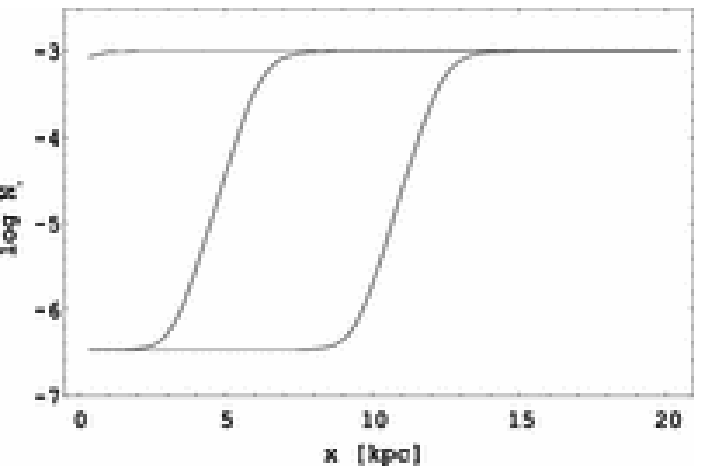}}
\caption{\label{unifden} The time evolution of the logarithm of the number 
density of neutral hydrogen $N_{0} (x,t)$  at three different 
times  $t= 0, 2 ,4$~ Ma in uniform density models. From top to bottom:
$N_{\rm tot}=10^{-5}$ cm$^{-3}$,~ $I_{\rm{irr}}=10^5$ erg~cm$^{-2}$
(Model a); $N_{\rm tot}=10^{-4}$ cm$^{-3}$,~ $I_{\rm{irr}}=10^5$ erg~cm$^{-2}$
 (Model b);
$N_{\rm tot}=10^{-3}$cm$^{-3}$,~$I_{\rm{irr}}=10^5$ erg~cm$^{-2}$ (Model c);
$N_{\rm tot}=10^{-3}$cm$^{-3}$,~
$I_{\rm irr}=3\times 10^5$ erg ~cm$^{-2}$ (Model d)}.
\end{figure}

To our knowledge there does not exist an analytical solution for equation 
\ref{rateeq} combined with equation  \ref{transfeq}, which is a system of non-linear
partial differential equations.
Therefore, we have to solve the discretized equations numerically. 
An immediate possibility is to employ the method of lines (cf. Schiesser, 1991)
which would incorporate an up-wind discretisation in the spatial coordinate and
the use of a solver for the resulting system of ordinary differential equations
(as e.g. a Stoer-Burlisch method).
Unfortunately, since the right hand sides of both the rate and the transfer equations are 
always negative in our cases, such an discretisation leads to instabilities 
whenever the temporal 
stepsize exceeds a certain value (cf. Kahaner et al., 1989).
Moreover, the system is often very stiff as a consequence of the large differences in
the time-scales associated with photoionisation and recombination. As a consequence,
in explicit schemes unacceptably short time steps have to be employed. 
Therefore, we employ here an implicit scheme, i.e. an up-wind discretisation in 
x-t space. Only first order differencing is considered here in order to avoid
overshooting.
In general, this involves the solution of a huge system of simultaneous 
non-linear equations (cf. Schiesser, 1991). Fortunately,
this is not necessary here, when the fact is exploited that light rays propagate 
always in one direction, i.e. the specific
intensity at a spatial point $x^{i+1}$ depends only on the specific intensity at 
$x^i$ and the absorption coefficients $\sigma N_0$ 
and the source function $S$ in $dx=x^i \dots x^{i+1}$. If we restrict ourselves 
to linear approximations these quantities
can be assumed to be given by $\sigma N_0(x^{i+1},t)$ and $S(x^{i+1},t)$.

The discretized equations \ref{rateeq} and \ref{transfeq} therefore read
\begin{eqnarray}
\label{discret1}
&&\frac{{\it N}_0(x^{i+1},t^{k+1})-{\it N}_0(x^{i+1},t^{k})}{t^{k+1}- t^k} 
\nonumber \\
&& \quad =- \sigma {\it N}_0(x^{i+1},t^{k+1}) I(x^{i+1},t^{k+1}) \nonumber \\
&& \quad+\alpha \left({\it N}_{tot}(x^{i+1})-
{\it N}_0(x^{i+1},t^{k+1}) \right)^2 
\end{eqnarray}
\begin{eqnarray}
\label{discret1a}
&&\frac{I(x^{i+1},t^{k+1})- I(x^{i},t^{k+1})}{x^{i+1}-x^i} \nonumber \\
&&+\frac{1}{c}\frac{I(x^{i+1},t^{k+1})-I(x^{i+1},t^{k})}{t^{k+1}-t^{k}} \nonumber \\
&& \quad= 
- \sigma {\it N}_0(x^{i+1},t^{k+1}) \nonumber \\
&& \quad \quad \times \left[ I(x^{i+1},t^{k+1})-S(x^{i+1},t^{k+1})\right]
\end{eqnarray}
or if we solve equation \ref{discret1a}
for $I(x^{i+1},t^{k+1}) $ and insert the result into Eq.  \ref{discret1} we get
\begin{eqnarray}
\label{discret2}
&&\frac{{\it N}_0(x^{i+1},t^{k+1})-{\it N}_0(x^{i+1},t^{k})}{t^{k+1}- t^k} 
\nonumber \\
&&  =- \sigma {\it N}_0(x^{i+1},t^{k+1}) 
\nonumber \\
&& \times\frac{1}{d}\left[
I(x^{i+1},t^{k})(x^{i+1}-x^{i})+I(x^{i},t^{k+1})c(t^{k+1}- t^k)
\right. \nonumber \\
&& \left.  +\sigma {\it N}_0(x^{i},t^{k+1}) c(t^{k+1}- t^k)(x^{i+1}-x^{i})
S(x^{i+1},t^{k+1}) \right]
\nonumber \\
&& +\alpha \left({\it N}_{tot}(x^{i+1})-
{\it N}_0(x^{i+1},t^{k+1}) \right)^2 \nonumber \\
\end{eqnarray}
and
\begin{eqnarray}
\label{discret2a}
&&I(x^{i+1},t^{k+1}) \nonumber \\
&& =\frac{1}{d}\left[I(x^{i+1},t^{k})(x^{i+1}-x^{i})+I(x^{i},t^{k+1})c(t^{k+1}- t^k)
\right. \nonumber \\
&& \left.  +\sigma {\it N}_0(x^{i},t^{k+1}) c(t^{k+1}- t^k)(x^{i+1}-x^{i})
S(x^{i+1},t^{k+1})\right] \nonumber \\
&&
\end{eqnarray}
with
\begin{eqnarray}
&&d=(x^{i+1}-x^{i})+c(t^{k+1}- t^k) \nonumber \\
&&\quad +(x^{i+1}-x^{i})c(t^{k+1}
- t^k)\sigma {\it N}_0(x^{i+1},t^{k+1})
\end{eqnarray}
Equation \ref{discret2} is a cubic equation  for ${\it 
N}_0(x^{i+1},t^{k+1})$,
i.e. it is an equation in one variable, whenever $ I(x^{i},t^{k+1})$ is known. 
Therefore, if
we make use of the boundary conditions, we can  march through the $x-t$ plane as 
indicated in
Figure \ref{grid_march}. The equation can be solved analytically or by a simple 
numerical solver. This is much
more economic in memory and CPU requirements than the general case, where one 
has to solve at
every time-step a non-linear system, in which each equation depends on the 
densities at all
spatial gridpoints. In this way we get an algorithm that is unconditionally 
stable for all
spatial and temporal stepsizes  and that is very efficient even for very many 
grid points.

We demonstrate that our method converges to the required steady states for 
Model b3 (see Table 1) in Fig. \ref{istcomp}. In order to calculate the steady state model
by an independent method we have first set $\partial N_0 / \partial t =0$ in the
rate equation, solved the algebraic equation for $N_0$ and inserted the result into the transfer equation
which we subsequently solved as an ordinary differential equation by means of a predictor-corrector
scheme.

Depending on the computer and operating system as well as the software used the CPU times 
range from a few seconds to a few minutes. This is about a factor $1000$ more efficient than
an explicit method of lines approach.

\subsection {The inclusion of many atomic levels}

If the  details of the ionising flux are to be taken into account, the 
relevant wavelengths $\lambda$ and the wavelength dependence of opacities 
and of the source function have to be considered explicitly. Therefore the 
rate equation and the transfer equation read 
\begin{eqnarray}
\label{rateeqn}
\frac{\partial{\it N}_0( x,t)}{\partial t} &=& -{\it N}_0({x},t)\int_0^{911\AA}  
\sigma(\lambda) I(x,t,\lambda)d\lambda \nonumber \\
&& \qquad +\alpha ({\it N_{\rm tot}}- {\it N}_0({ x},t))^2
\end{eqnarray}

\begin{eqnarray}
\label{transfeqn}
&& 
\frac{1}{c}\frac{\partial I(x,t,\lambda)}{\partial t}+
\frac{\partial I(x,t,\lambda)}{\partial x} 
\nonumber \\
&&\qquad =-\sigma(\lambda){\it 
N_0}(x,t)\left( I(x,t,\lambda)
-S(x,t,\lambda)\right) \nonumber \\
&&
\end{eqnarray}

The discretized versions (cf. equations \ref{discret2}) can now be written

\begin{eqnarray}
\label{discretn}
&&\frac{{\it N}_0(x^{i+1},t^{k+1})-{\it N}_0(x^{i+1},t^{k})}{t^{k+1}- t^k} 
\nonumber \\
&&  =-  {\it N}_0(x^{i+1},t^{k+1}) \sum_{l=1}^{L} w^l \sigma(\lambda^l) 
\nonumber \\
&& \times\left[
\frac{1}{d(\lambda^l)}\left(I(x^{i+1},t^{k},\lambda^l)(x^{i+1}-x^{i})
\right. \right. \nonumber \\
&&\left. \left. +I(x^{i},t^{k+1},\lambda^l)c(t^{k+1}- t^k)
\right. \right. \nonumber \\
&& \left.\left.  +\sigma(\lambda^l) {\it N}_0(x^{i},t^{k+1}) c(t^{k+1}- t^k)(x^{i+1}-x^{i})
\right. \right. \nonumber \\
&&\left. \left.
\times S(x^{i+1},t^{k+1},\lambda^l)\right)
\right] \nonumber \\
&& \quad+\alpha \left({\it N}_{tot}(x^{i+1})-
{\it N}_0(x^{i+1},t^{k+1}) \right)^2 \nonumber \\
\end{eqnarray}
with
\begin{eqnarray}
&&d(\lambda^l)=(x^{i+1}-x^{i})+c(t^{k+1}- t^k) \nonumber \\
&&\quad +(x^{i+1}-x^{i})c(t^{k+1}
- t^k)\sigma(\lambda^l) {\it N}_0(x^{i+1},t^{k+1})
\end{eqnarray}
($w^l$ are the weights for the wavelength integration). It is seen that the 
first equation contains only
${\it N}_0(x^{i+1},t^{k+1})$  as an unknown. However, the polynomial for the 
determination of ${\it N}_0(x^{i+1},t^{k+1})$
is now of degree $2L+1$. Fortunately, this does not slow down the 
calculations significantly if a numerical solver is used.

\subsection{Generalisation to more than one source of radiation}

In the general case, we need to consider the presence of more than one source of 
radiation. We describe the algorithm we use in terms of two sources of
radiation, but the generalisation to many sources is immediate.
 
Suppose we have two sources of illumination: one is situated at $x=0$ and
shines from time $t=0$ to time $t=t_1$ with constant luminosity and the other one is placed at
$x=x_{end}$ and shines from $t=t_2$ to $t=t_{end}$ where the subscript `end' indicates
the maximum value. We now employ the two-stream-approximation where $I_p$ indicates the direction
of increasing $x$ values and $I_m$ the opposite direction. The mean intensity is then given by
$J=(I_p+I_m)/2$. Corresponding neutral hydrogen densities are denoted by $N_{0p}$ and $N_{0m}$. 
The solution is derived now by a fixed-point iteration for $N_0(x,t)$. 
For the first step $I_m^1(x,t)=0$ is assumed for all $x$ and $t$ values. $I_p^1(x,t)$ and 
$N_{0p}^1(x,t)$ are then calculated for all spatial and temporal grid points  as described above.
Next, $I_m^1(x,t)$ and $N_{0m}^1(x,t)$ are calculated starting from $t=0$ and $x=x_{end}$ (using the same
 scheme as above but just
going in the opposite direction) with the $I_p^1(x,t)$ as just derived. $N_0^1(x,t)$ is then calculated as the 
algebraic mean of $N_{0p}^1(x,t)$ and $N_{0m}^1(x,t)$.   
Next, $I_p^2(x,t)$ and $N_{0p}^2(x,t)$ are evaluated with the assumption that the intensity in the opposite
direction is given by the values of the previous iteration, and then $I_m^2$ and $N_{0m}^2(x,t)$ leading to 
$N_0^2(x,t)$. The next iteration then follows, and we stop the process when
 $\vert N_0^m(x,t)-N_0^{m-1}(x,t) \vert / N_0^m(x,t) < \varepsilon$
for all spatial and temporal grid points.  For $\varepsilon=10^{-5}$ this 
was always the case for $m \leq 5$ in our computations,
i.e. we find a fast convergence.

For the models discussed in this paper, the $\frac{1}{c}\frac{\partial 
I(x,t)}{\partial t}$ term is unimportant and is therefore not included. 
The role played by this term is discussed in section 3.3. Since we are
not dealing with recombination radiation in this paper, we set $S(x,t)=0$ 
in our calculations. Furthermore, unless 
explicitly stated, all our illustrative models assume two levels.

\section{Model results for single sources}

\subsection{Plane parallel uniform density slabs}

We assume that the medium is plane parallel extending from $x=0$ to $x=x_{\rm max}$. 
The density profile is specified to be

\begin{eqnarray}
\label{densform} 
N_{\rm tot}(x,0) = N_{\rm{amb}}+N_{\rm{enh}} \exp[- (\frac{x-x_s}{l_w})^2]~\rm{cm}^{-3}
\end{eqnarray}
where  $N_{\rm{amb}}$ is the ambient density. The second term allows for a density 
enhancement 
with a Gaussian profile of amplitude $N_{\rm{enh}}$ and half width $l_w$ centered at 
$x=x_s$.  
 
We choose values for the free parameters to correspond approximately to the 
conditions that 
are expected in the  universe when the first sources of radiation turned on.  
For the standard
 cosmological parameters, the gas density at redshift $z$ is given by 
$N_{\rm tot}\sim 10^{-7} (1+z)^3$ so we consider a density range  $10^{-5} -10^{-3}$ 
cm$^{-3}$ appropriate to 
$z\sim 6 - 20$. We set the density enhancements to peak at $\sim 1$ dex 
above the mean. 

We place the source of UV photons at $x=0$ and obtain results
for a  spatial region that extends  to $x_{\rm max} = 20$ kpc. 
The inflow intensity at the outer boundary of the plane parallel region that is 
being irradiated 
will depend on the luminosity of the source, its geometry,  and its location 
relative to the 
region. In our examples we use an incident intensity
$I_{irr}=10^5$ erg~cm$^{-2}$ s$^{-1}$ which corresponds to the total intensity
of a black body of temperature $T=10^5$ K (a UV source) diluted by a 
factor $\sim 10^{-10}$. Since these are 1-D models, we do not specify the
nature of the sources or of the dilution.

\begin{figure*}
\center
\begin{minipage}{140mm}
\scalebox{0.9}{\includegraphics{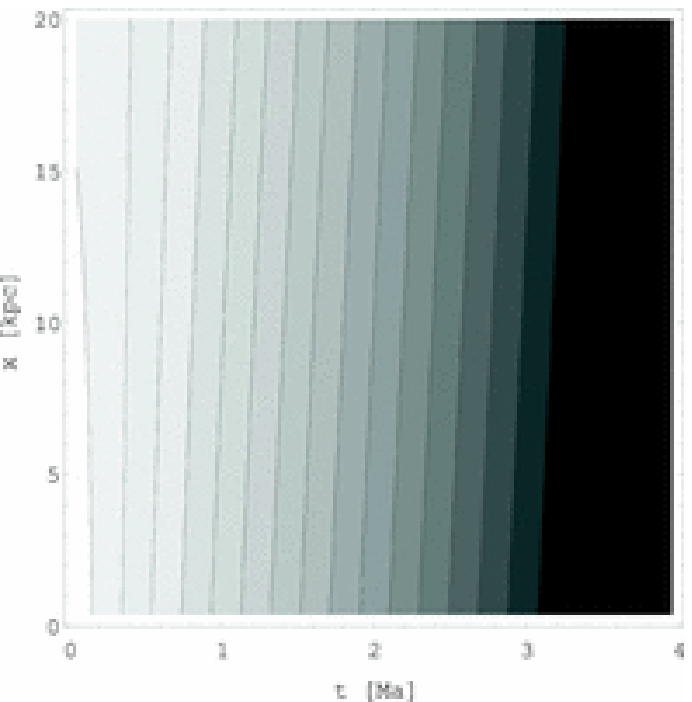}}
\hspace{1mm}
\scalebox{0.9}{\includegraphics{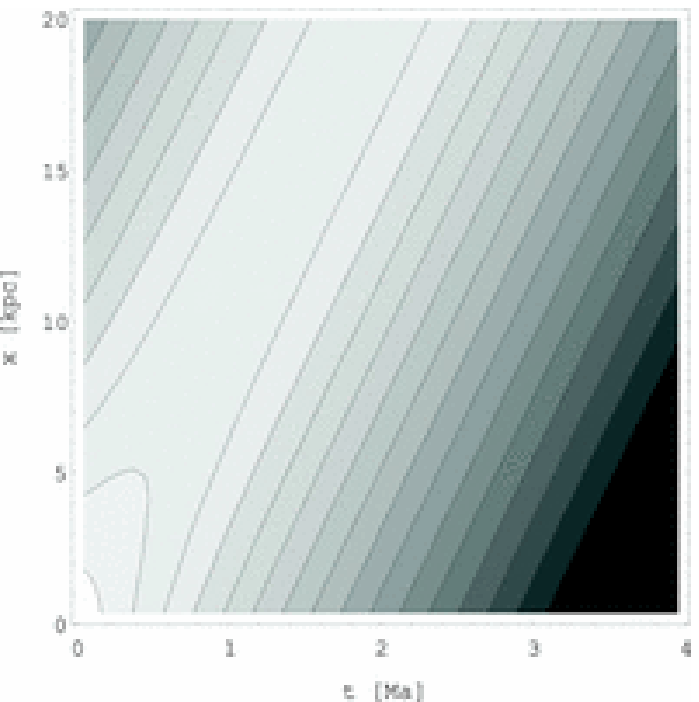}}
\vspace*{0.4cm}

\scalebox{0.9}{\includegraphics{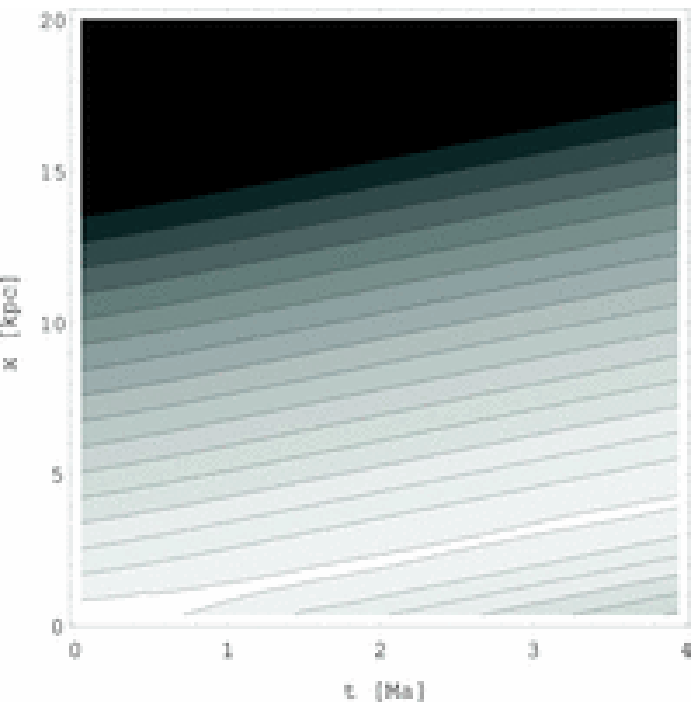}}
\hspace{1mm}
\scalebox{0.9}{\includegraphics{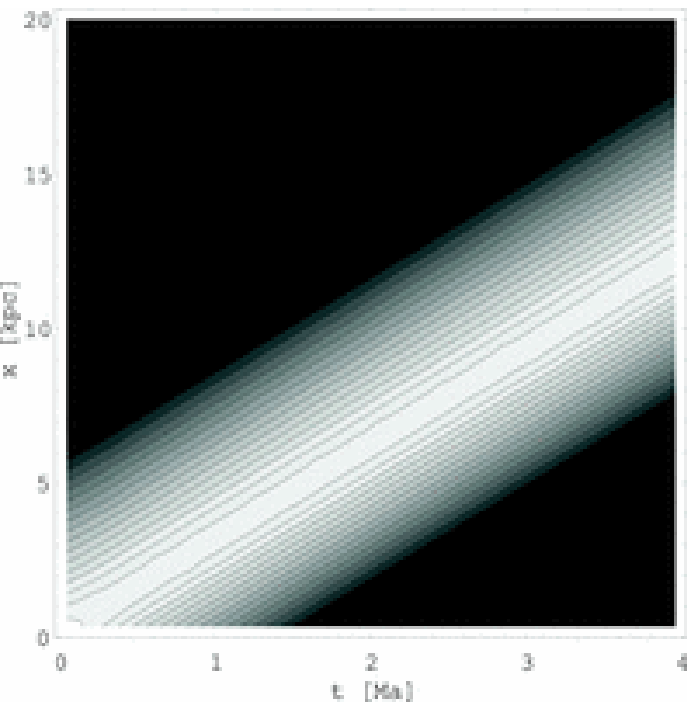}}

\end{minipage}
\caption{\label{unifden_fronts} The rate of change of the neutral density 
$\vert\frac{dN_{0}}{dt}\vert$ on a 
logarithmic scale for 
Models a (top left),b (top right),c (bottom left) and d (bottom right). The front corresponds to 
the region of highest gradient
shown here with the lightest shading. The speed of the front is seen to decrease 
as the density increases for a given incident intensity, and to increase 
as the incident intensity increases at a given density.}
\end{figure*}

Equation  \ref {rateeq}  can be used to estimate a recombination time scale
\begin{eqnarray}
\label{eqnrecomb1}
t_{\rm rec}\sim \frac{N_0}{\alpha(N_{\rm tot}-N_0)^2}
\end{eqnarray}
Initially, when the gas is mainly neutral, recombination is ineffective 
($t_{\rm rec} \rightarrow \infty$),
 and the time evolution is 
governed by photoionisation. As the degree of ionisation increases, $t_{\rm rec}$ 
decreases, and for a gas 
that is  $50\%$ ionised,  the above expression yields a characteristic 
recombination time scale (using $
\alpha \sim  10^{-18}$ cm$^2$ (Osterbrock (1974))
\begin{eqnarray}
\label{eqnrecomb2}
t_{\rm rec} \sim 6.3 \times 10^{2} ~[\frac {10^{-5} {\rm cm}^{-3}} {N_{\rm tot}}] ~{\rm 
Ma}
\end{eqnarray}
This time scale becomes smaller, the higher the density.

Likewise, using $\sigma \sim 10^{-18}$ \rm cm$^2$ (Osterbrock 1974) equation 
\ref{transfeq} yields a photo-ionisation time scale
\begin{eqnarray}
\label{eqnphotion}
t_{\rm phot} \sim \frac{1}{\sigma I} \sim 0.32~ [\frac{10^5 {\rm erg~ 
cm}^{-2}s^{-1}}{I}] ~{\rm Ma}
\end{eqnarray}
This time scale is smallest for the highest intensities. As  the intensity of 
radiation decreases (due to photo-absorption), 
the photo ionisation time  scale increases, with  $t_{\rm phot} \rightarrow  \infty$ 
as $I \rightarrow 0$. 
In principle, the radiation field and the ionised fraction can adjust 
so as to reach equilibrium at all points in the medium. However, in the 
situations that we consider the level populations never 
reach full equilibrium during the lifetime of the source.  Our calculations, 
however, allow us to chart the rate at which this 
equilibrium is approached at different times at different points in the medium.
 
In the first set of calculations we do not allow for a density enhancement 
($N_{\rm{enh}}=0$) and consider three models with different 
uniform densities $N_{\rm{amb}} = 10^{-5} $~cm$^{-3}$ (Model a), $N_{\rm{amb}} = 10^{-4} 
$~cm$^{-3}$ (Model b) and $N_{\rm{amb}} = 10^{-3} 
$~cm$^{-3}$ (Model c). Figure \ref{unifden} shows the run of density 
of neutral hydrogen for these three cases at 
three different times $t= 0, 2, 4$ Ma. For the model with the 
lowest ambient density (Model a), the entire 
computational region is optically thin to UV radiation (even at  $t=0$) so that 
the effects of radiative transfer are minimal. 
As a consequence, the density of neutral gas is nearly independent of $x$, but 
decreases with time.  The decline occurs exponentially 
on  the photoionisation time scale being  dominated by the first term in the 
rate equation \ref{rateeq}. The neutral  component does 
not show any marked spatial structure at a given time within the domain of 
calculation up to the time $t_{\rm max}$ 
when the calculation is terminated.

As the ambient density is increased by a factor $10$ (Model b; Figure 
\ref{unifden}) optical depth effects become more important.  
Regions that are closest to the source are least affected by transfer effects  
and ionise more quickly in comparison to regions 
that are further away. As a consequence recombinations become relatively more 
important 
as one moves away from the source. This is shown by the gradual increase in the 
neutral gas density with increase in $x$  
at any given time $t >0$. In this model there are clear changes in 
$\frac{dN_{0}}{dt}$ which -- as  
we shall see -- can be used to define the location of the ionisation front.

As the ambient density is increased even further (Model c; Figure \ref{unifden}),
 optical depth effects dominate except very close 
to the source. A well defined ionisation front which propagates outwards with 
time is now clearly seen within our computational domain. 
The effect of increasing the incident intensity by a factor of three to 
$I_{irr} = 3\times 10^5$ erg cm$^{-2}$ s$^{-1}$ for the same density distribution 
is shown in Model d. The ionsiation front is now seen to propagate further 
in the same time interval.

Comparing the  models in Figure \ref{unifden}, we see that the maximum 
degree of ionisation that is achieved at $t=t_{\rm max}$ ranges from $10^{-5}$ to  
$10^{-3}$ which  
agrees in order of magnitude with the mean observed degree of ionisation
of the 
diffuse intergalactic medium at the tail end of re-ionisation (Fan et al. 2003).

The properties of the fronts are best illustrated in Figure \ref{unifden_fronts} 
where we present 
$\vert\frac{dN_{0}}{dt}\vert$ as a function of $x$ and $t$ for Models a, b ,c
and d. We identify the 
front as the region of the maximum rate of change of ionisation (in an absolute 
sense). 
These diagrams clearly show how the front propagates in time and changes its 
spatial structure. 
The properties of the fronts, such as the speed, can be extracted from this data 
and these are
given in Table 1. 
 
In Model a, the front propagates fast as it eats its way through the low density 
medium. The speed 
of the front is highest for this model.  In Model b, 
the higher density slows down the front, so that it advances at a slower rate
while in Model c, the even higher density reduces the speed further. We note
that in these calculations, the front speed is essentially independent of time 
except close to $t=0$ where the speed is not well defined (see section 3.3). 

The front speed is a strong function of the irradiating intensity as shown in Figure \ref{unifden_fronts}. In general, a 
stronger source will result in a front that propagates faster qualitatively 
similar to a  model with a lower mean density.


\begin{figure}
\scalebox{0.8}{\includegraphics{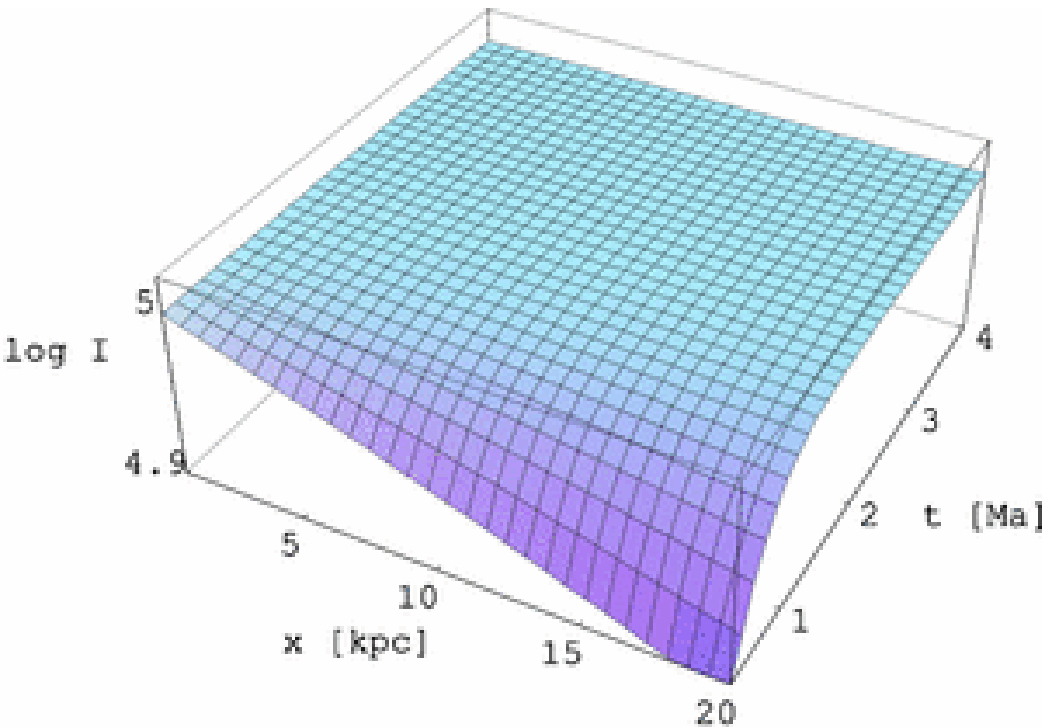}}
\scalebox{0.8}{\includegraphics{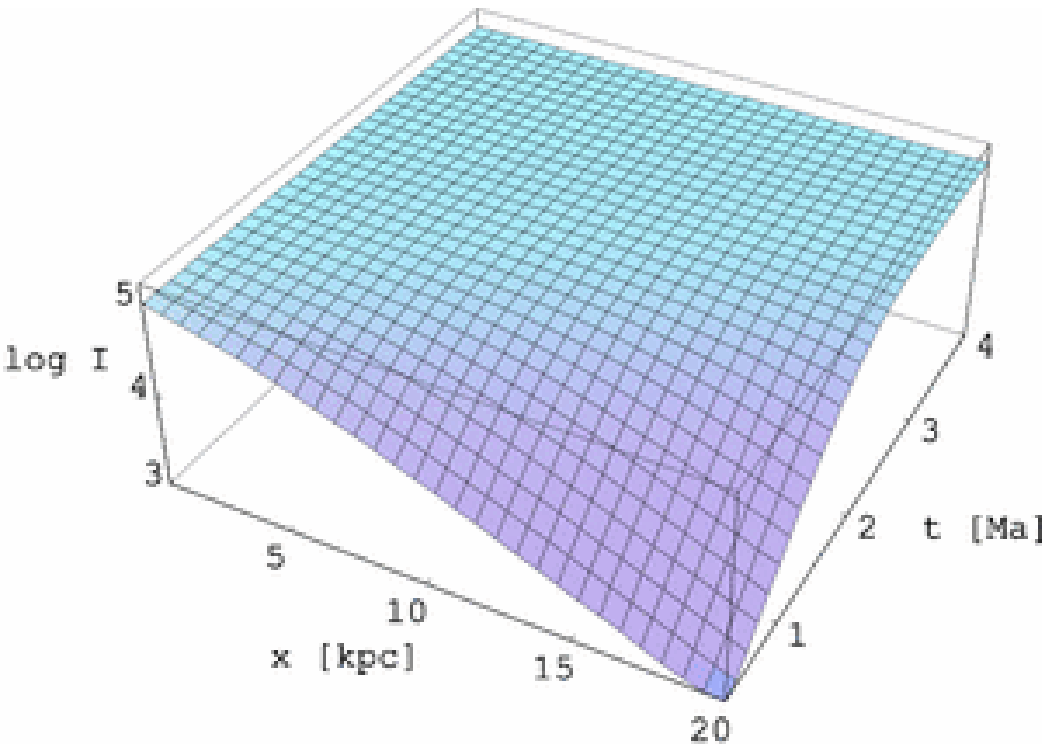}}
\scalebox{0.8}{\includegraphics{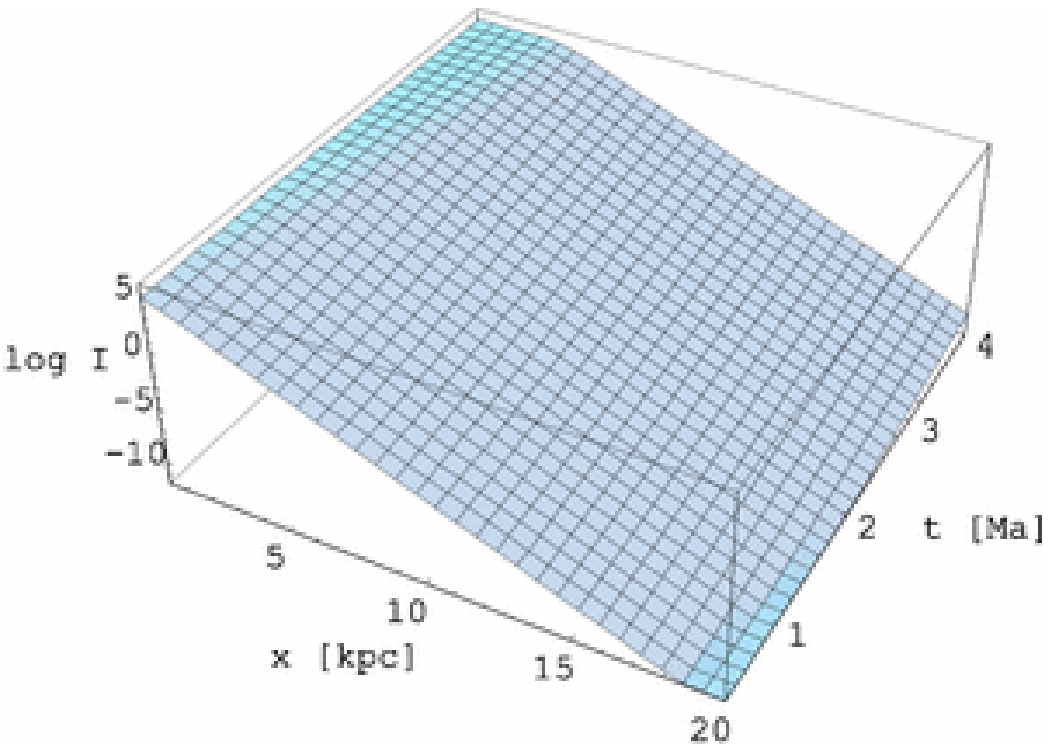}}
\caption{\label{unifden_I} The spatial distribution and time evolution of the
 logarithm of the intensity of radiation 
of Models a (top left), b (top right) and c. These calculations show how
 the source emerges
from the dark phase to become bright and visible as the front propagates through 
the computational
volume}
\end{figure}

\begin{figure}
\scalebox{0.35}{\includegraphics{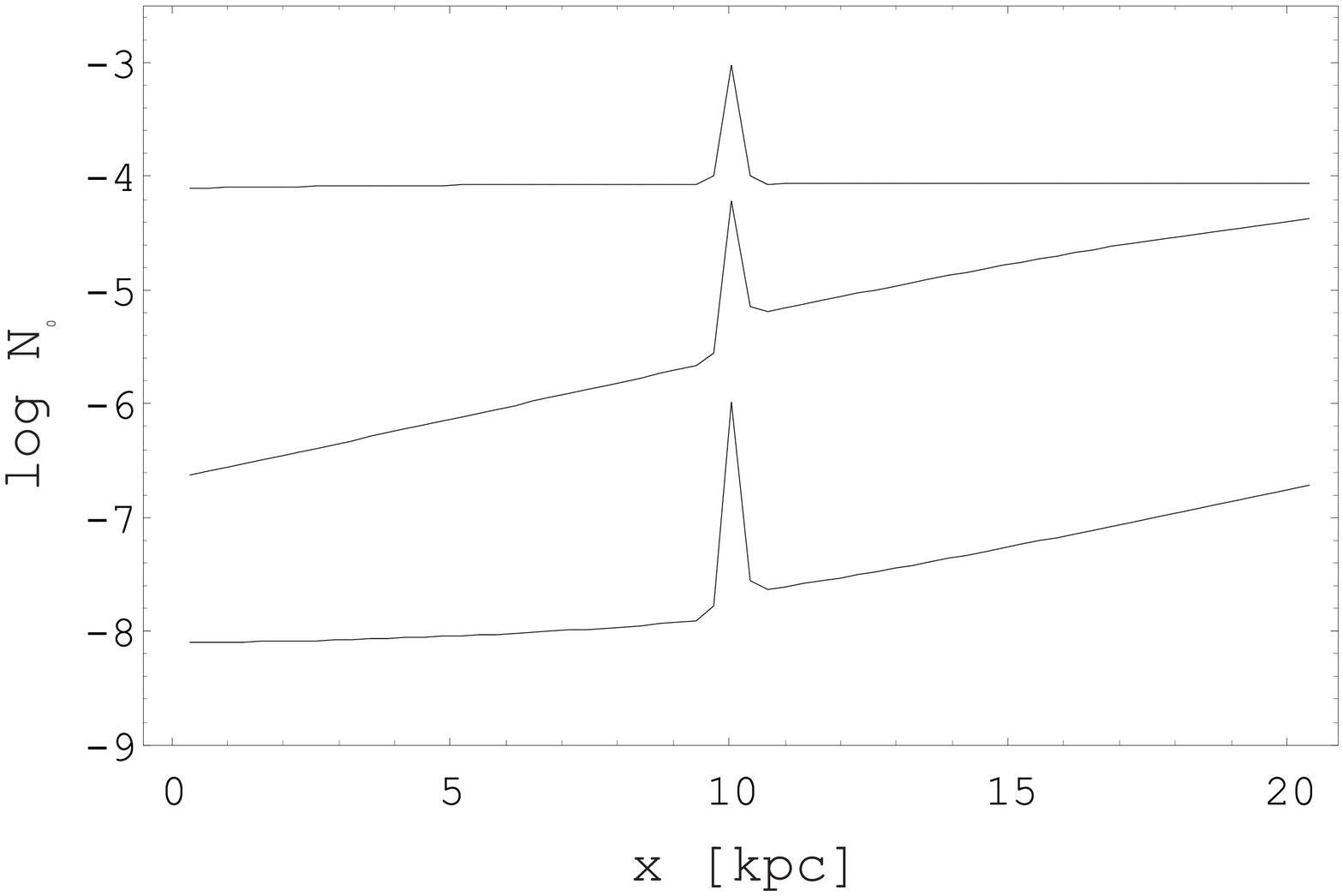}}
\vspace*{0.4cm}

\scalebox{0.35}{\includegraphics{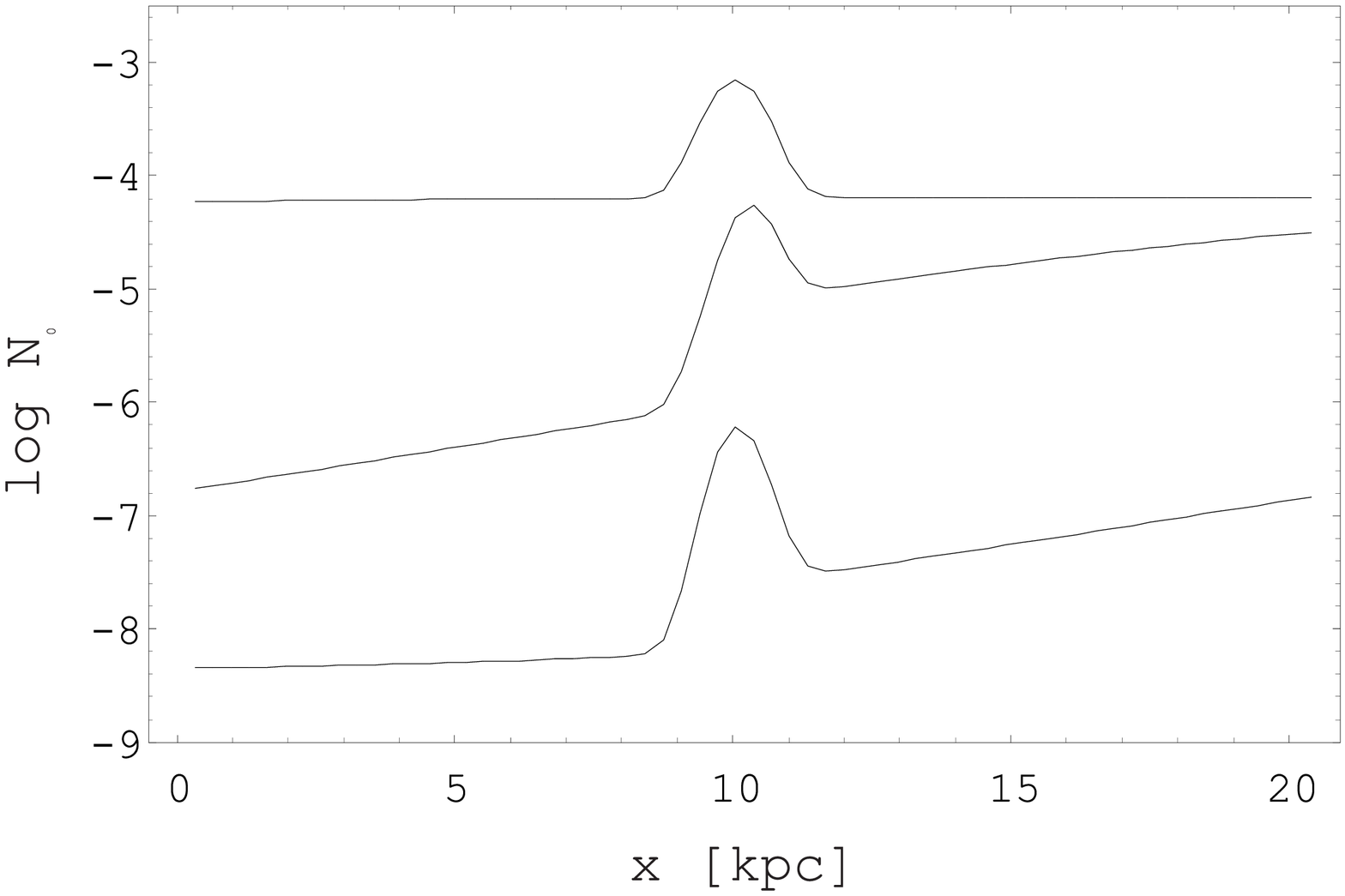}}
\vspace*{0.4cm}

\scalebox{0.35}{\includegraphics{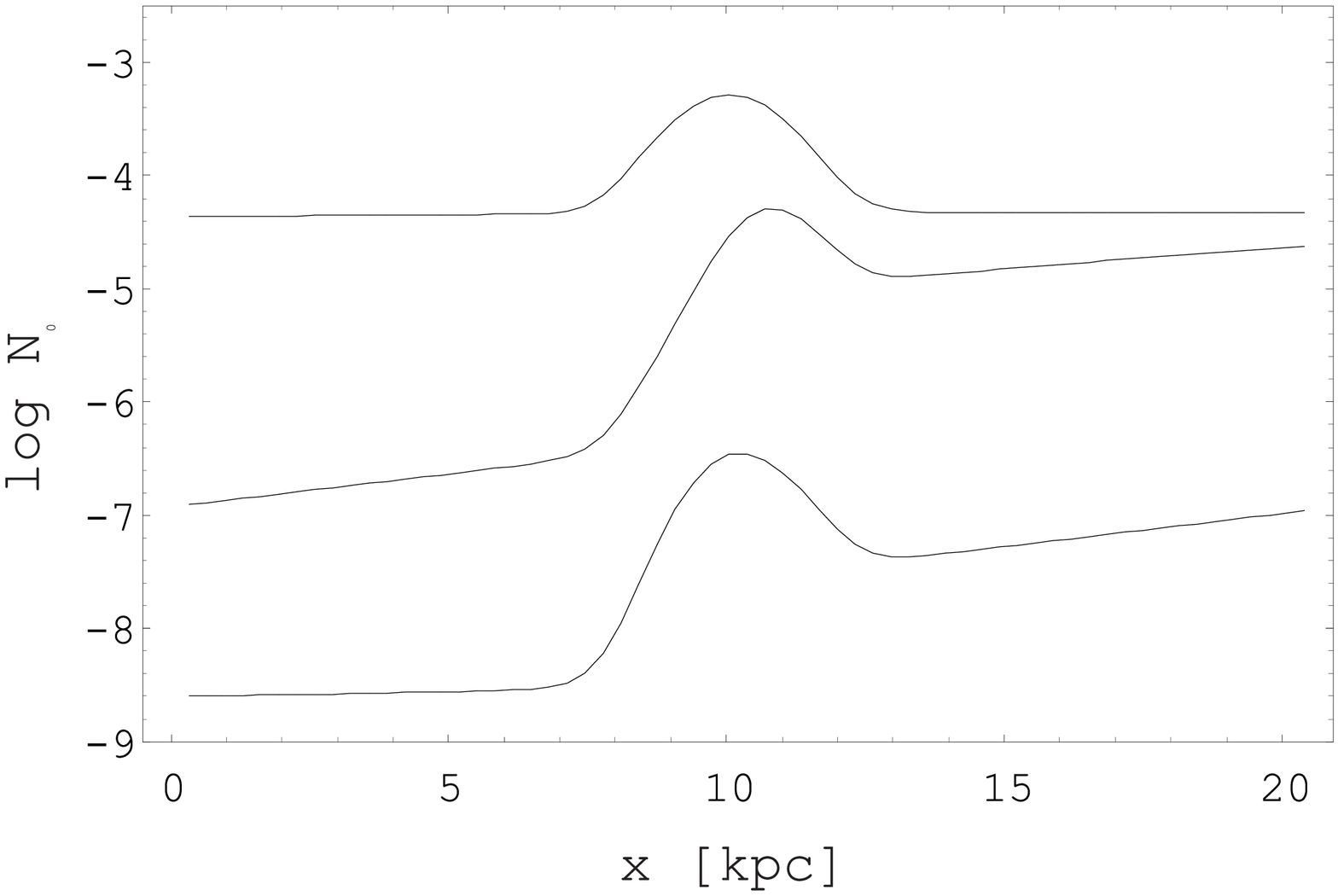}}
\vspace*{0.4cm}

\scalebox{0.3}{\includegraphics{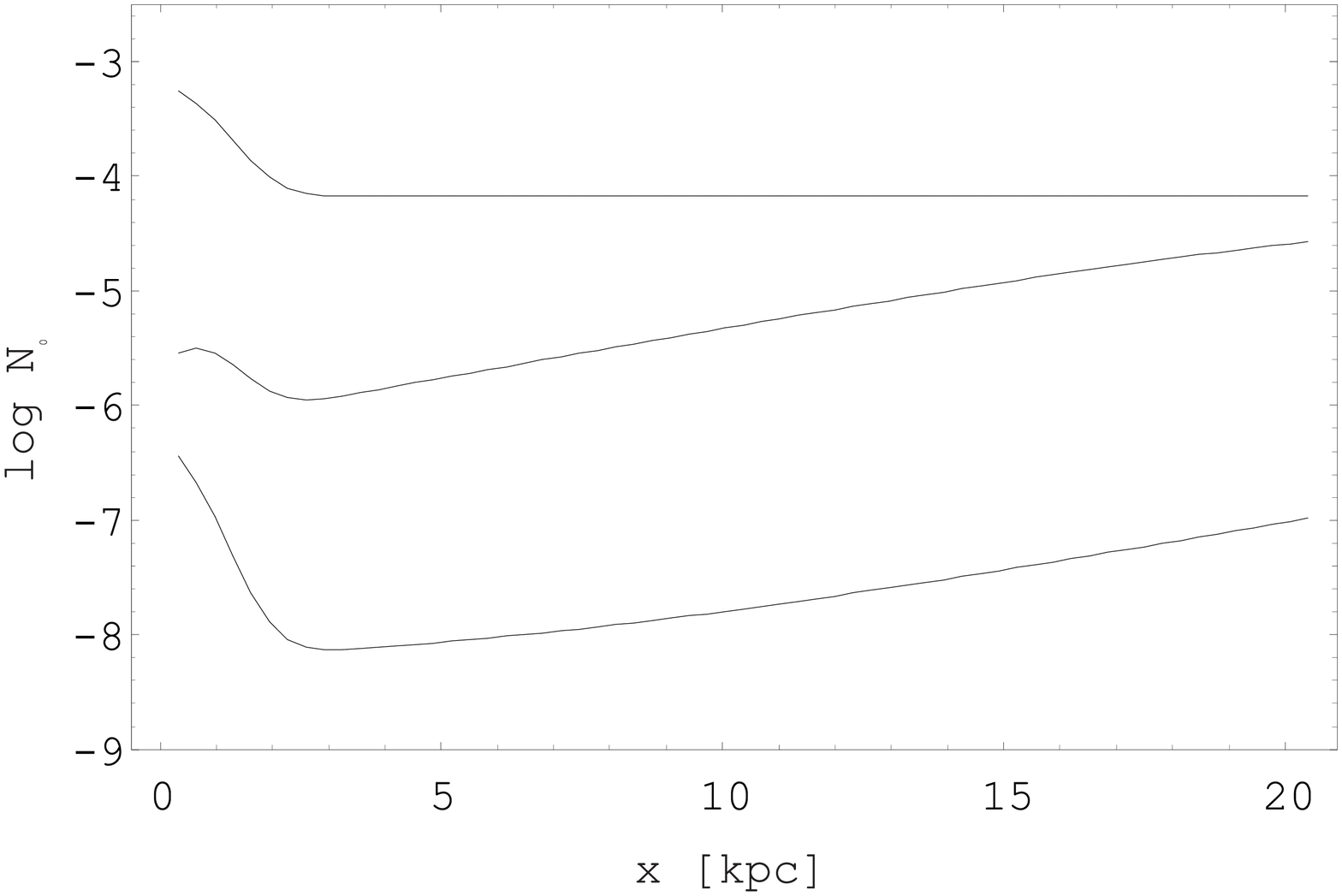}}
\caption{\label{nonunif} 
The time evolution of the logarithm of the number density of neutral hydrogen 
$N_{0} (x,t)$  at
three different times  $t= 0, 2, 4$~Ma in a non uniform medium with 
density
enhancements described by Gaussian profiles. The parameters of the four models 
(Models b1, b2, b3, b4 from top to bottom) are given 
in Table 1. All models are normalised to have a total column density 
$\Sigma = 6.4 \times 10^{18} $ cm$^{-2}$ corresponding to Model b}
\end{figure}

The spatial distribution and time evolution of the intensity of radiation for 
Models a,b and c are 
shown in Figure \ref {unifden_I}. As already noted, this sequence 
corresponds to one of increasing optical depth at time $t=0$ when the gas is 
neutral. Very close to
the source, the intensity is maintained at essentially the input intensity at 
all times considered due 
to the dominance of photoionisation. The intensity that emerges 
from the computational region (i.e. at  $x=x_{\rm max}$), on the other hand, evolves 
with time at 
different rates depending on the density of the model. In  Figure \ref 
{unifden_I} (Model a), 
the emergent intensity is essentially the input intensity with little change as 
$t$ approaches
$t=t_{\rm max}$ since photoionisation dominates over the entire computational 
volume.  On the other hand, 
in Figure \ref {unifden_I} (Model b), this intensity increases from a low value at 
initial times to a value close to 
the input intensity as  $t $ approaches  $t_{\rm max}$ and a larger fraction of the 
gas becomes ionised
and therefore less opaque. The source will therefore first appear dimmed when 
viewed from $x=x_{\rm max}$ brightening as $t$ increases towards $t_{\rm max}$. In a 
sense, this source 
emerges from a dark phase (the  `dark ages' in a cosmological context) within 
the maximum time
used in our computations. In contrast, at the even higher densities represented 
in  
the model in Figure \ref {unifden_I} (Model c), the bulk of the 
material remains neutral for all $t \le t_{\rm max}$ so that the source never 
emerges from the 
dark phase. Clearly, had these calculations been carried out for even larger 
times, the
ionisation front will eventually move outwards to $x_{\rm max}$ and the source will 
no longer 
be dark. We note that apart from a scaling factor, the inverse of the intensity 
in 
Figure \ref {unifden_I} is proportional to the photoionisation time 
scale.

\subsection{Plane parallel slabs with Gaussian density enhancements}

In the second set of models we introduce a density enhancement in the middle
of the computational 
region at $x_s=10$ kpc (Models b1, b2 and b3) and at $x_s=0$ (Model b4) 
parametrised by $N_{enh}$ and $l_w$. The widths of the enhancements 
increase from Model b1 through to 
Model b3, and the densities are 
normalised so that the total column number density is always $\Sigma = 6.4 \times 
10^{18} $ cm$^{-2}$. 
The parameters for this series of models are also summarised in Table 1.

We show in Figure \ref{nonunif} a, b, c and d  the density of neutral hydrogen for
Model b1, b2, b3 and b4 at three different times $t= 0, 2, 4 $~Ma.

\begin{figure*}
\center
\begin{minipage}{140mm}
\scalebox{0.9}{\includegraphics{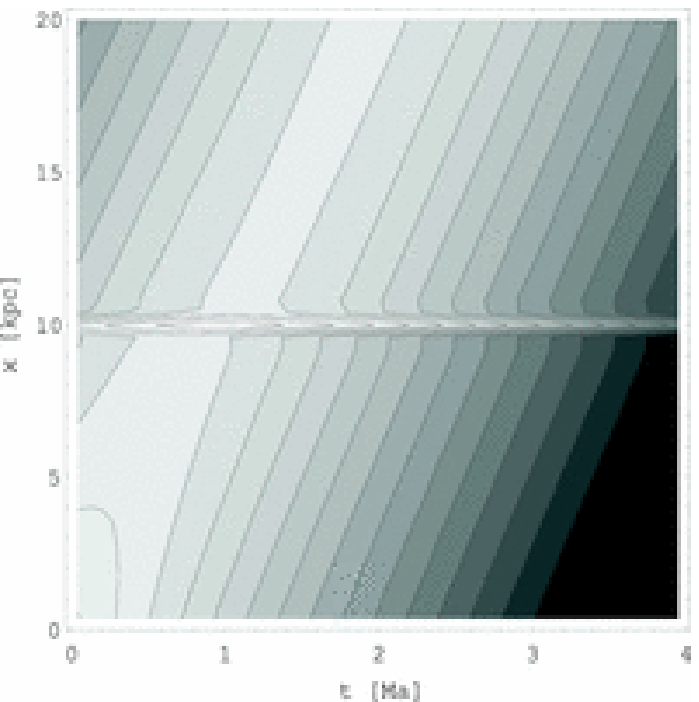}}
\scalebox{0.9}{\includegraphics{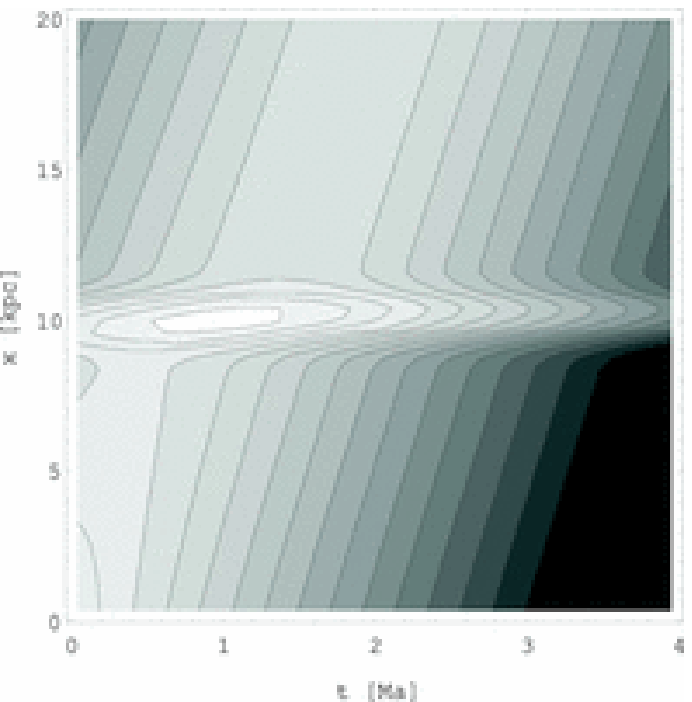}}
\vspace*{0.4cm}

\scalebox{0.9}{\includegraphics{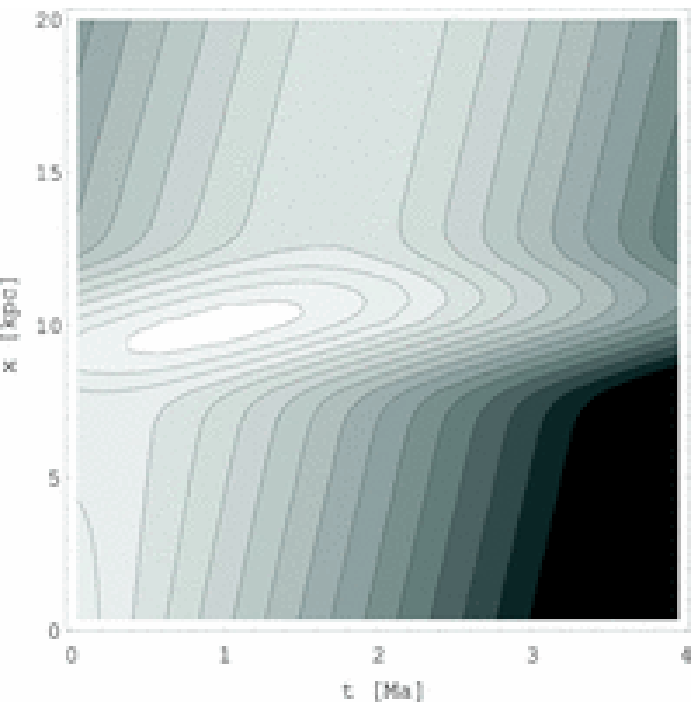}}
\scalebox{0.9}{\includegraphics{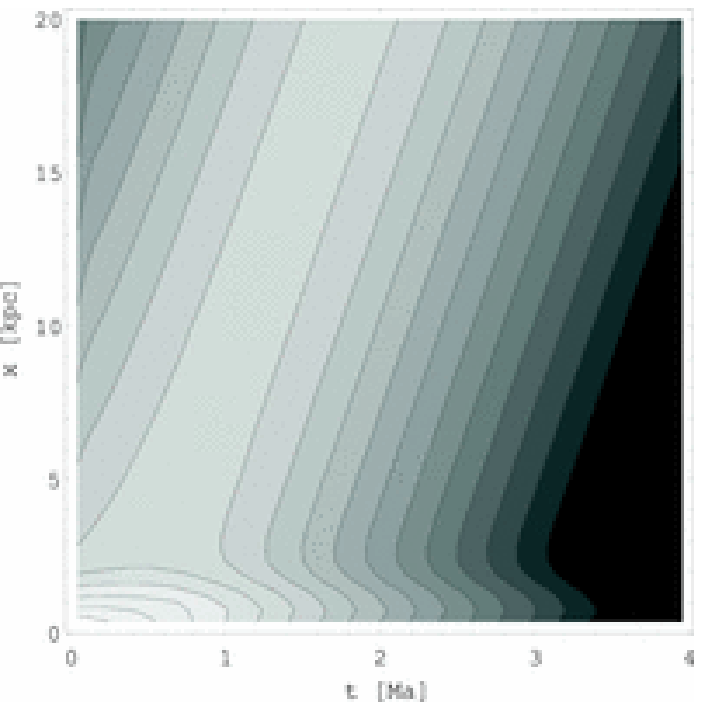}}
\caption{\label{nonunif_fronts}
The rate of change of the neutral density $\vert \frac{dN_{0}}{dt}\vert$ on a 
logarithmic scale
for the models with density enhancements; Model a (top left), Model b (top right), Model c (bottom left),
and Model d (bottom right)
showing the propagation of the ionisation fronts through a Gaussian density 
enhancement.
The speed of the front is reduced and a time delay introduced due to the 
encounter.}
\end{minipage}
\end{figure*}

\begin{figure*}
\center
\begin{minipage}{140mm}
\scalebox{0.7}{\includegraphics{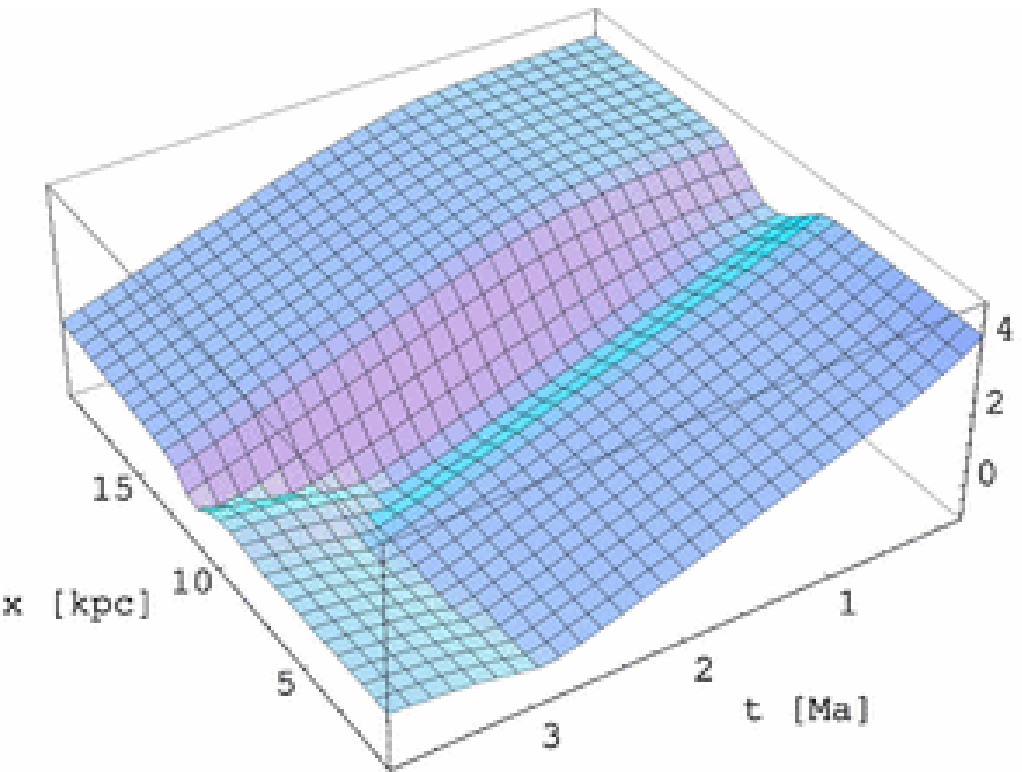}}
\scalebox{0.7}{\includegraphics{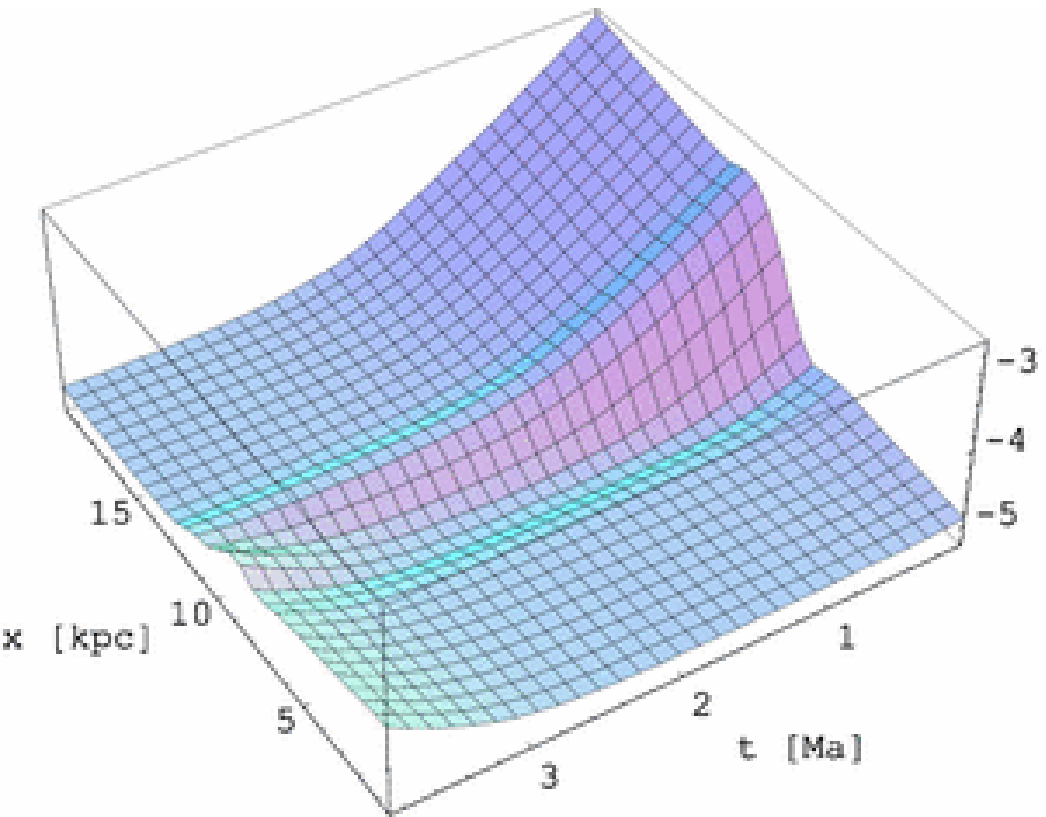}}
\caption{\label{eqmrate} The spatial and time dependence of the recombination 
time
scale (left) and the total time scale ${N_0}/\vert{\frac{dN_{0}}{dt}}\vert$ 
(right) 
for Model b3 which has 
a Gaussian density enhancement at $x=10$kpc}
\end{minipage}
\end{figure*}

The introduction of a density enhancement leads to a local increase in the extinction coefficient 
and therefore a decrease in the number of 
ionising photons that can propagate to 
the far side of the enhancement in Models a, b and c. As a result ionisation is more effective on 
the `near side' 
of the density
enhancement (the side that faces the source of irradiation) than on the `far 
side' as can 
be seen by the drop in the level  of ionisation as one crosses the enhancement.  
The model
with the largest value of $l_w$ (Model b3) shows the strongest drop in the level 
of ionisation 
at any given time. As $t$ increases, the overall level of ionisation increases 
and the optical depth through the density enhancement decreases. The relative change in the level of 
ionisation across the enhancement also decreases.

In Model b4, the source is at the peak of the density enhancement
so that the radiation propagates through a density distribution that
 decreases with distance from 
the source. The time evolution is very similar to what is found for the 
far side of the density enhancements in Models a, b and c. The ionisation 
proceeds slowly first close to the source, and then propagates outwards at
later times.

The interaction of the  outward propagating front with a density enhancement is 
beautifully
illustrated in Figure \ref{nonunif_fronts} which is analogous to Figure  
\ref{unifden_fronts}.
Here we see that as a front  with a velocity $v_1$ encounters  a density 
enhancement, 
it slows down dramatically to a mean speed $v_2$ as it adjusts to the higher 
density environment.  Eventually the front penetrates the density enhancement 
and emerges
at a speed $v_3$. $v_1$ and $v_3$ correspond very approximately to the speeds 
appropriate
to the  ambient density and intensity on the near and far sides of the density 
enhancement.
However, in general $v_3$ is larger than $v_1$ with the difference being largest 
for the enhancement with the largest Gaussian half width. The encounter creates a 
well defined delay $\Delta t_f$ in the process of ionisation which is a function of the 
parameters of the Gaussian.

In the case of Model b4, where the source is at the peak of the density
enhancement, the velocity is initially low, and then steepens very quickly 
as the front breaks through and the velocity approaches the value 
appropriate to the ambient density.

The time delays and the other properties of the fronts are summarised in
Table 1.
 
The above calculations show that in the presence of Gaussian density 
enhancements, even of 
modest amplitude, can result  in significant delays in the onset of 
re-ionisation. For suitable
choices of parameters (as in Model b3) that may be appropriate to the first 
sources of radiation 
that form in the early universe, the delay could exceed the lifetime of the 
source. Once the source dies out, the gas would recombine on the 
much larger recombination time scale $t_{\rm rec}$. Pockets of ionised regions would 
thus survive enshadowed by high density regions until another source of UV photons 
turns on in its vicinity and proceedes to ionise its surroundings.

\begin{figure}
\scalebox{0.7}{\includegraphics{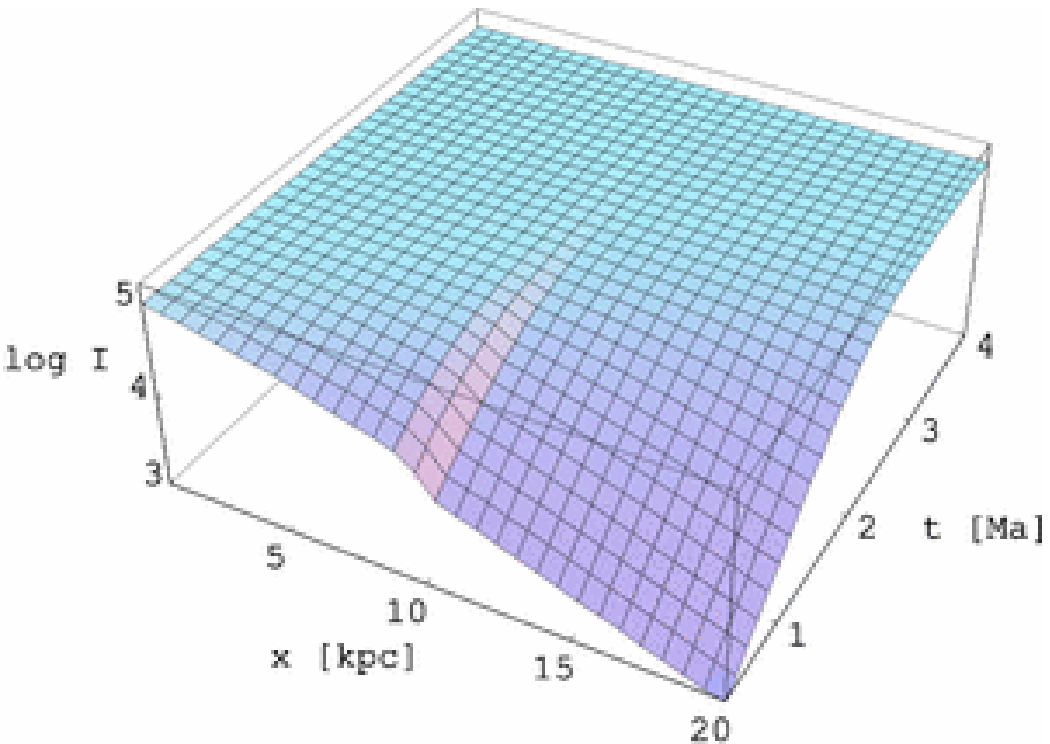}}
\scalebox{0.7}{\includegraphics{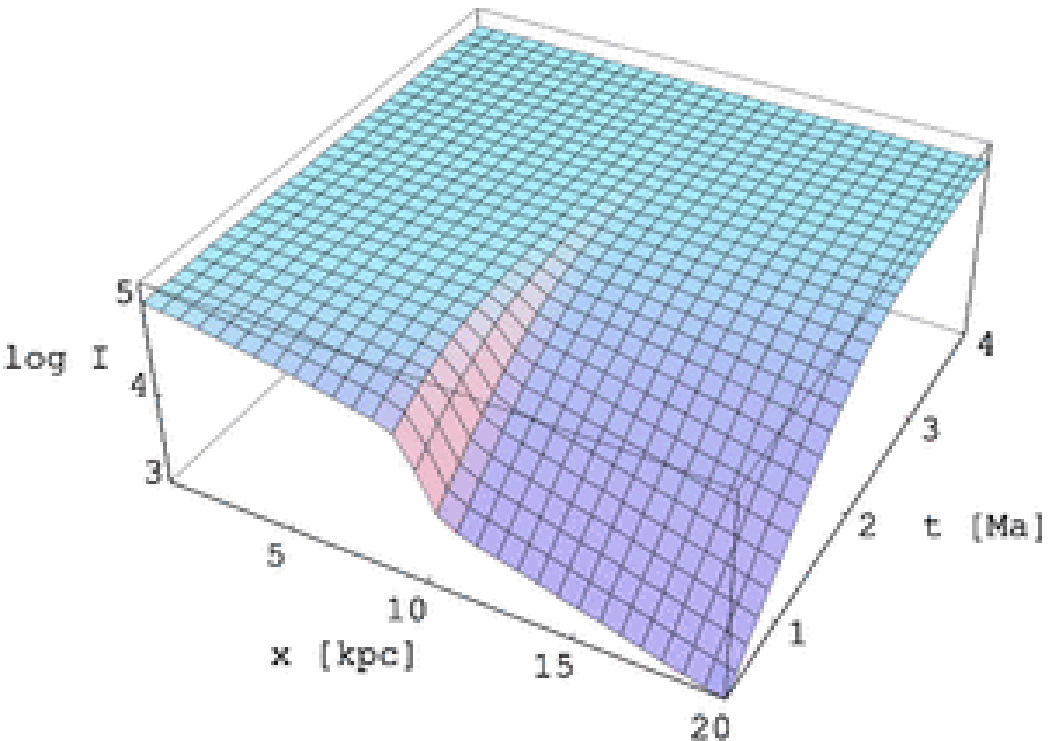}}
\scalebox{0.7}{\includegraphics{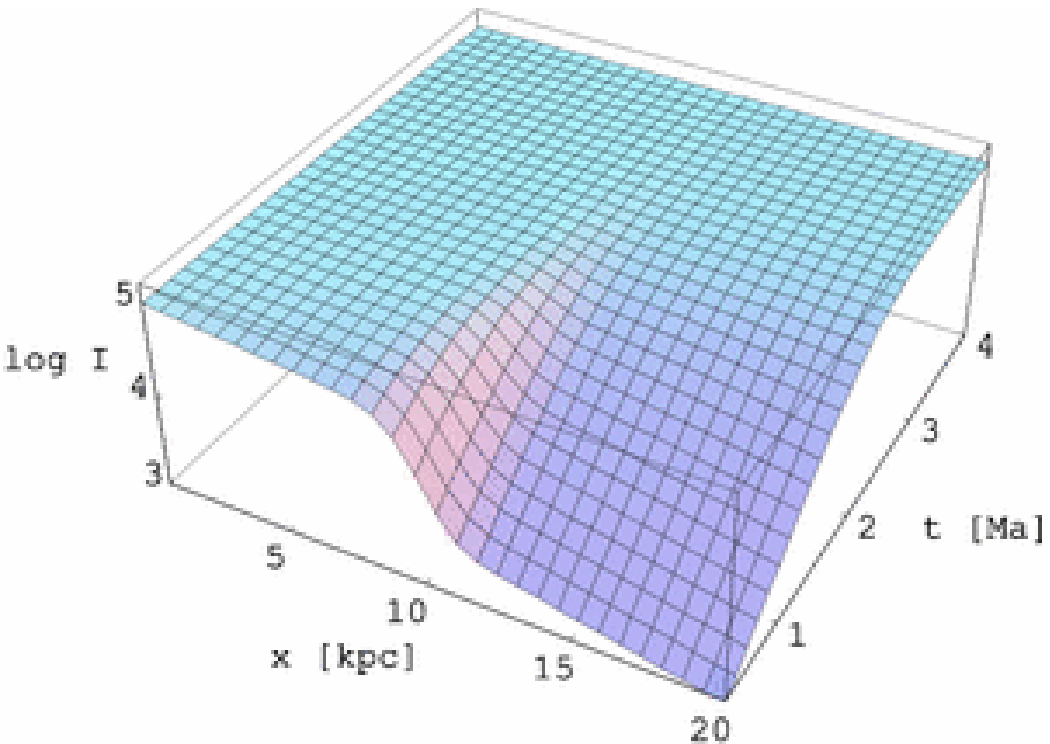}}
\caption{
\label{nonunif_I}
The spatial and time evolution of the logarithm of the intensity of radiation
for the models in Figure \ref{nonunif}. Apart for a scaling factor, the inverse 
of the intensity 
is proportional to the photoionisation time scale}
\end{figure}

The spatial distribution and time evolution of the intensity of radiation  for 
the Models
b1, b2 and b3 are shown in Figure \ref{nonunif_I}. These 
figures should 
be compared with the corresponding uniform density  Model b in Figure 
\ref{unifden_I}  which
has the same total column number density. The intensity is seen to drop sharply 
across the density
enhancement initially due to the local increase in optical depth, but with time, 
as the region ionises 
and becomes less opaque, the intensity approaches the input intensity. 

We conclude this section by presenting calculations of the recombination time 
scale and the actual
time scale on which the number of neutral hydrogen atoms change in Model b3 
(Figure \ref{eqmrate})
as a function of $(x,t)$.  The diagram shows that initially the recombination 
time scale has a
minimum at the peak of the Gaussian 
density enhancement. Comparing with the actual time scale we see that the 
photoionisation time scale 
is relatively more important close to the source than further away due to 
shielding effect of the density enhancement.  As time increases the shielding 
effect decreases and  recombination time scale becomes more uniform in the 
spatial domain. We note that these time scales can vary over several orders of 
magnitude 
across the ionising region so that it is not possible to describe the time 
dependent evolution 
using average recombination and photoionisation time scales.

\subsection{The role of time dependent term}

\begin{figure}
\scalebox{0.9}{\includegraphics{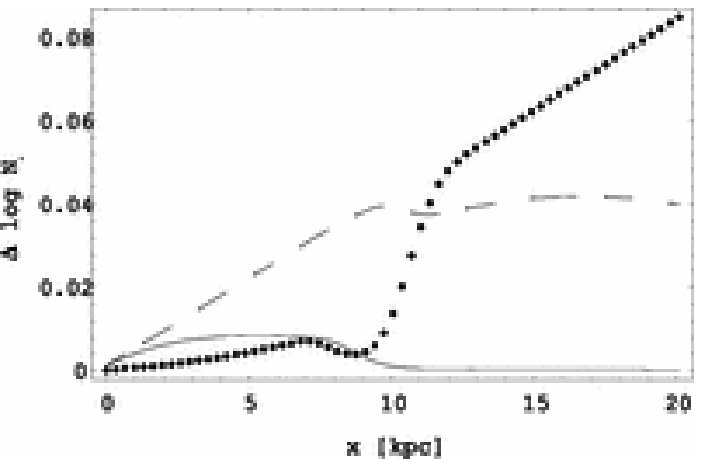}}
\caption{
\label{timedep}
The spatial and time evolution of the logarithm of the difference between the
calculated neutral density  with and without the term $\frac{1}{c} \partial I
/\partial t$ in the transfer equation for Model b3 (see Figure \ref{nonunif}).
The solid line refers to $.03$ Ma, the broken curve to $2$ Ma and the 
dotted curve to $4$ Ma. The light travel time across the computational
domain is $6.5\times 10^{-3}$~Ma.}
\end{figure}

The calculations that we have presented so far have neglected the time dependent 
term
$\frac{1}{c}\frac{\partial I(x,t)}{\partial t}$ in the transfer equation. To 
illustrate
the role played by this term, we have repeated the calculations for our model 
Model b3 
discussed in the previous section 
including this term. 
The results are presented in Figure \ref{timedep}.

The term with the time derivative in the transfer equation \ref{transfeq} 
may be 
expected to play an important role when $x >> ct$. The neglect of this
term could 
 result in faster than light propagation of ionising fronts. The light travel 
time across our computational region is $6.5\times 10^{-3}$~Ma, and  
in our calculations, 
this
condition is satisfied only at very early times which we have not attempted 
to resolve. The results of Figure \ref{timedep}, however,
show that when taken in conjunction with the rate equations, this term has an 
additional 
effect on the state populations at all spatial locations yielding values that 
are
lower than when the term is neglected.
The effect arises from the non-linearity of the rate equations, and its 
magnitude depends on  spatial location and time. Although we have neglected this term in our present models, our 
method allows for its inclusion with little additional cost in computing time.
This term may play a role in cosmological calculations. 

The approach to the steady state solution of this particular model 
was shown in Figure \ref{istcomp}. It is clear that none of the models 
in Figure \ref{nonunif} have reached a steady state in 
$t_{\rm max} \sim 4$~Ma, the assumed lifetime of the source indicating 
that full time dependent calculations are required to estimate 
the scale of ionisation domains.

\subsection {The global evolution of ionisation with time}

\begin{figure}
\scalebox{0.3}{\includegraphics{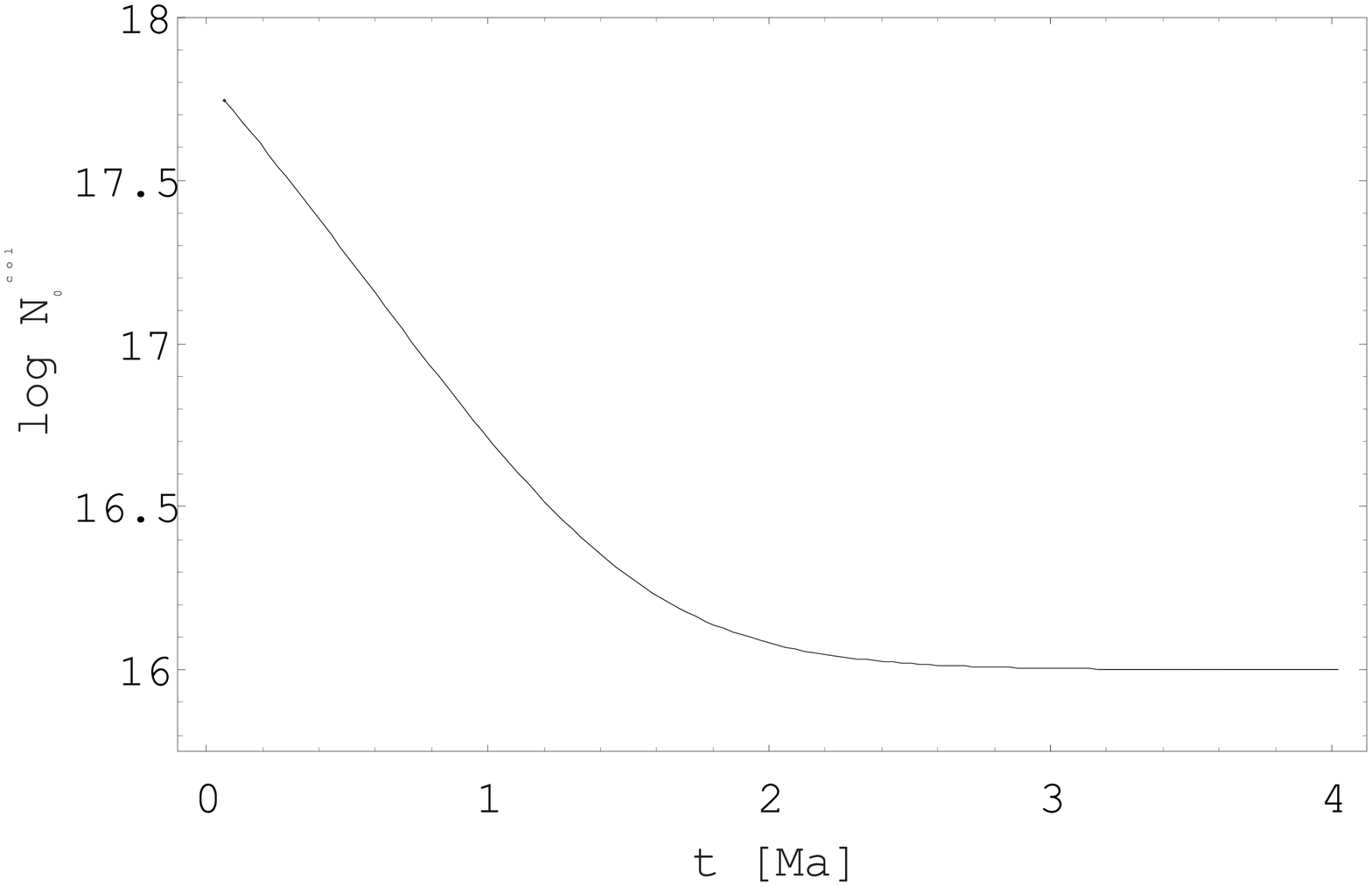}}
\scalebox{0.3}{\includegraphics{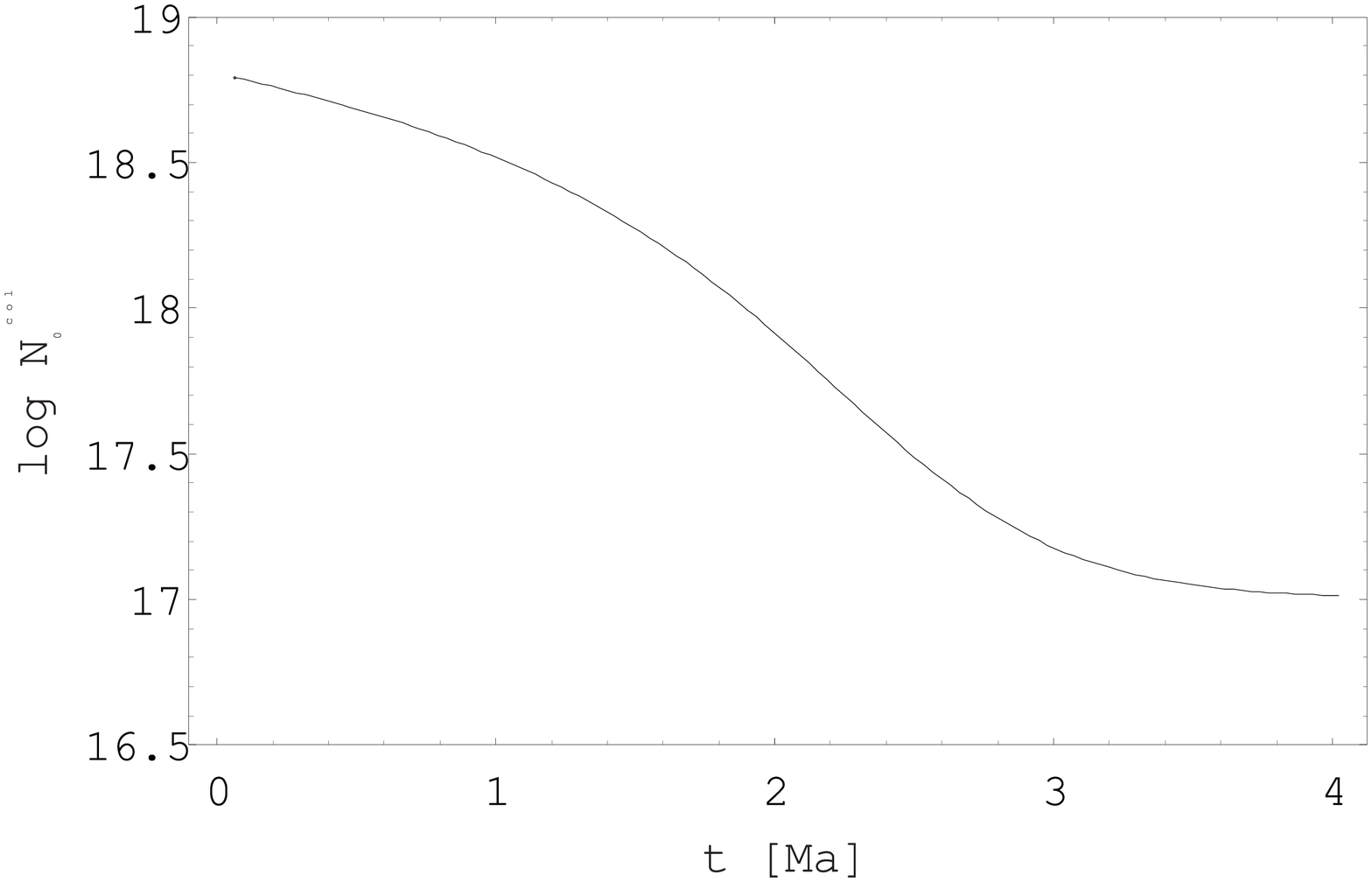}}
\scalebox{0.3}{\includegraphics{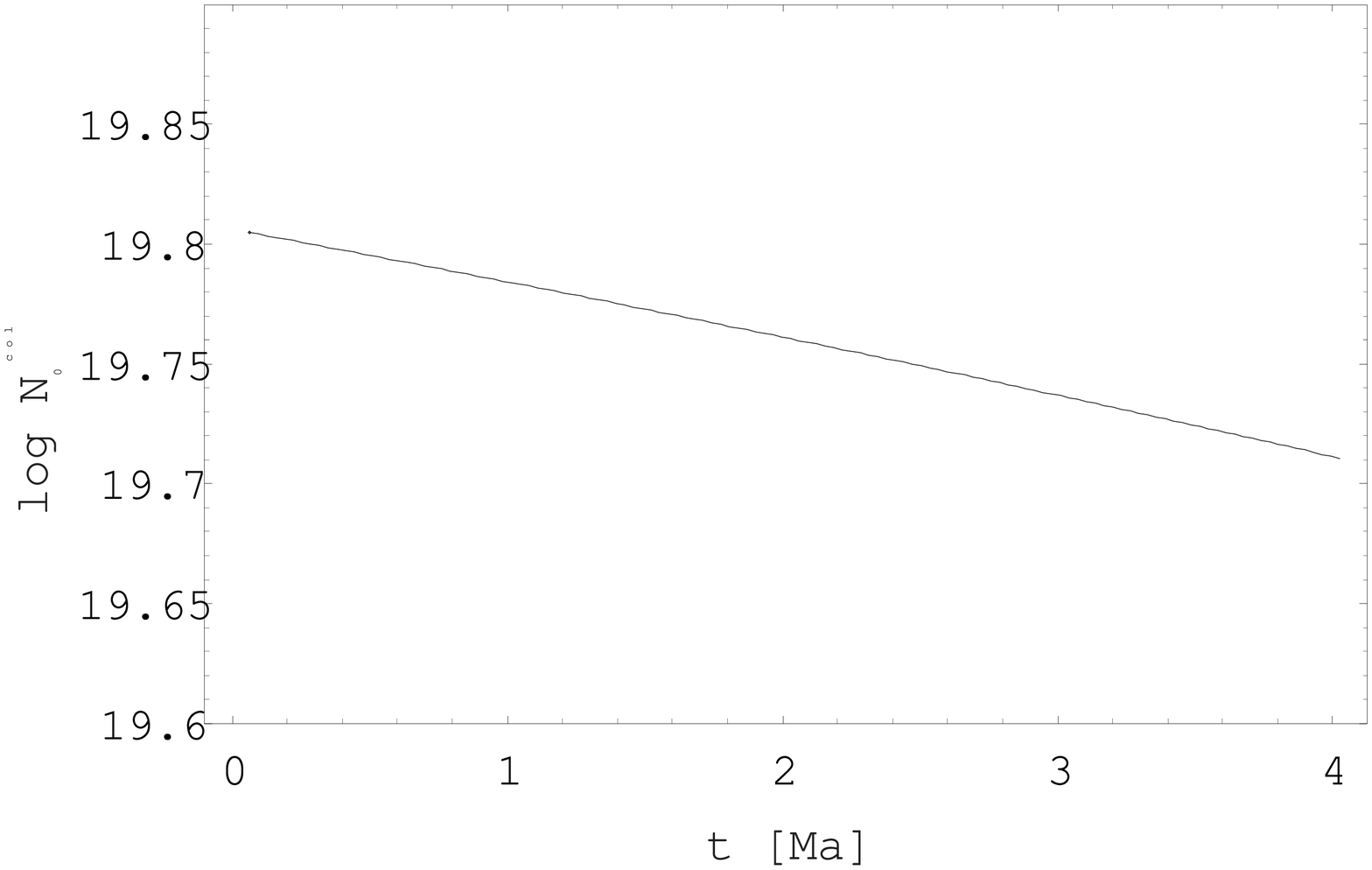}}
\caption{
\label{ion_m}
The time evolution of the logarithm of the neutral gas density within the
 entire 
computational volume for the uniform density Models a (bottom), b (middle)
and c (top). This figure shows the effects of increasing the value of 
$\frac{t_{\rm rec}}{t_{\rm phot}}$ by a factor $\sim 100$.
}
\end{figure}

A good global indicator of the time evolution of ionisation can be obtained by 
plotting
the number of neutral hydrogen atoms in our fixed computational volume as a 
function of time $t$. This is shown in Figure \ref{ion_m} for
the three uniform density models that have been discussed. The time evolution 
of the optical depths of these configurations can also be obtained from 
these diagrams by simply shifting the vertical scale by 
$\log \frac{1}{\sigma}\sim 18$.

In the very low density model (Figure \ref {ion_m}), the column density of 
neutral gas  
within the computational region is essentially constant decreasing only very 
slowly with time. For this model, the simple estimates based on equations \ref{eqnrecomb2} 
and \ref{eqnphotion}  give 
$t_{\rm rec}\sim 630 $~Ma,~$t_{\rm phot}\sim 0.32$~Ma. 
As the density is increased by a factor $10$ and optical depth effects 
become more important, the column density of neutral gas is seen to decline more 
rapidly initially due to photoionisation, but then more slowly 
as recombination begins to
play a significant role. For this model $t_{\rm rec}\sim 63 $~Ma.
These effects are seen even more strongly in the highest density model considered 
for which $t_{\rm rec}\sim 0.63 $~Ma.

We have not shown the corresponding curves for the models with density 
enhancements
discussed in Figures \ref{nonunif} since they are identical to the 
curve with $N_{\rm{amb}} = 10^{-4} $~cm$^{-3}$ in 
Figure \ref{ion_m}.  This is because all these four models have the same 
total column density in the computational region. The main effect of the
introduction of the density enhancement is to change the spatial distribution of ionised gas, 
and the time evolution of the ionisation fronts, but not the total column 
density of ionised gas.

The column density of ionised gas, and its evolution with time is therefore 
a strong function of the total column density and the incident radiation field 
and is not sensitive to the distribution of mass within the computational
region.

\begin{figure}
\scalebox{0.3}{\includegraphics{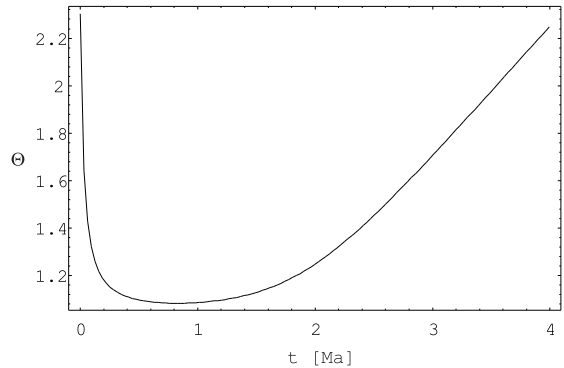}}
\caption{
\label{ratio}
The time evolution of the logarithm of the ratio $\theta(t)$ of the 
total number of ionising photons to the total
number of ionised particles 
in our computational box for Model b4.
}
\end{figure}

A related quantity that can be derived from our calculations is the 
time evolution of the ratio of the total number of ionising photons in our 
computational region to the total number of ionised particles. We have 
derived this 
quantity

$$
\theta(t) =\frac{\frac{4\pi}{c}\frac{1}{h\nu}\int_0^t
  I(x,t^\prime)dt^\prime}{\int N_e(x,t)dx}$$
for our model with a central density enhancement (Model b4) and the results 
are shown in Figures \ref {ratio}. 

Initially, the gas is neutral, and we  therefore expect $\theta$ to be large. 
As time proceeds, photons are removed from
this region generating ionised particles. As photoionisations proceeed,
$\theta$ declines rapidly, reaches a plateau, and rises again as the competing
effect of recombination comes into play.

\subsection{Effects of increasing the number of continuum states}
The models that we have presented so far have been based on a model atom 
with one bound state and one continuum state. To allow for  different spectral
slopes for the source of ionising photons and a better description of the
photoionisation cross section we need to include more than one 
continuum state.  
 
\begin{figure}
\scalebox{0.3}{\includegraphics{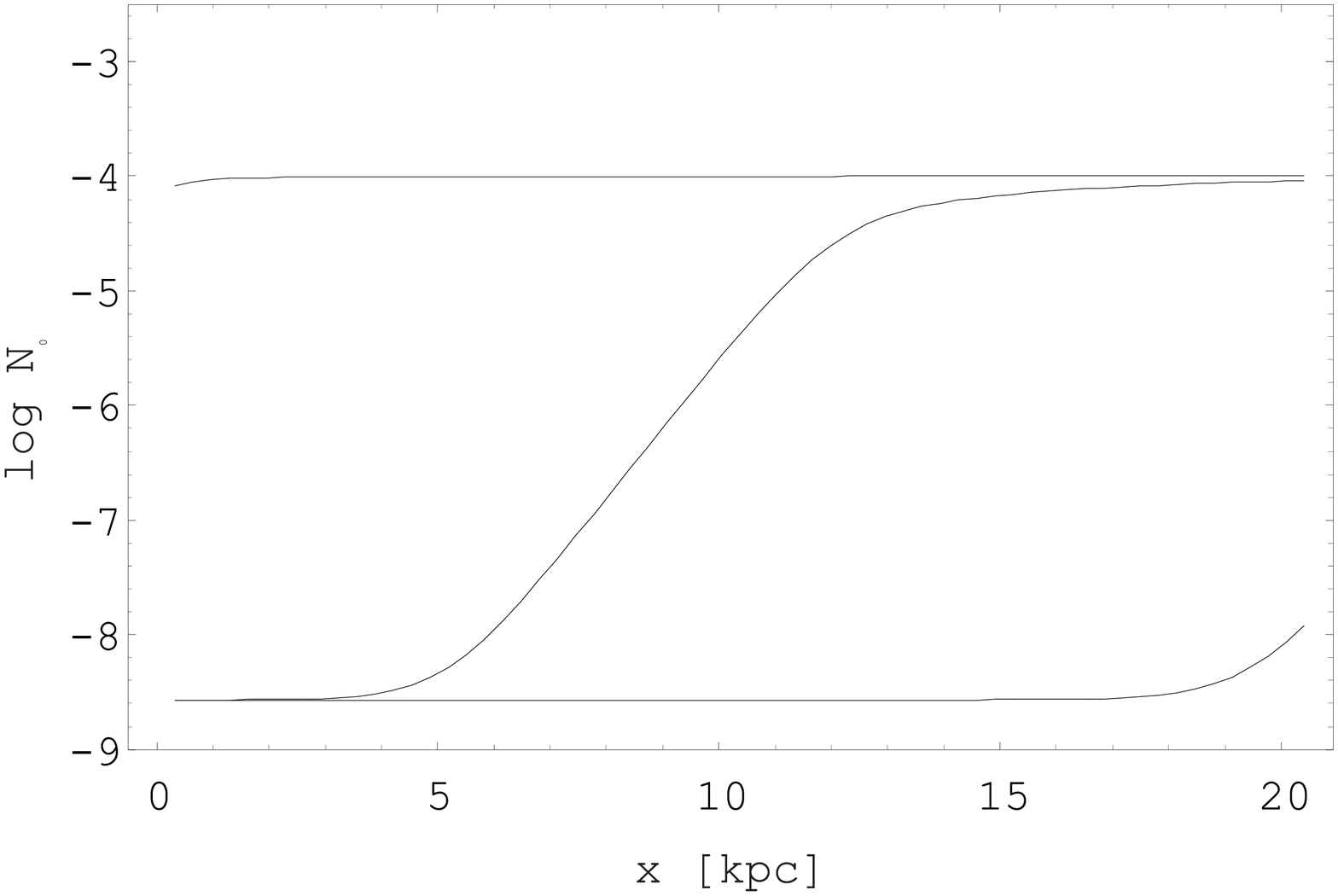}}
\scalebox{0.7}{\includegraphics{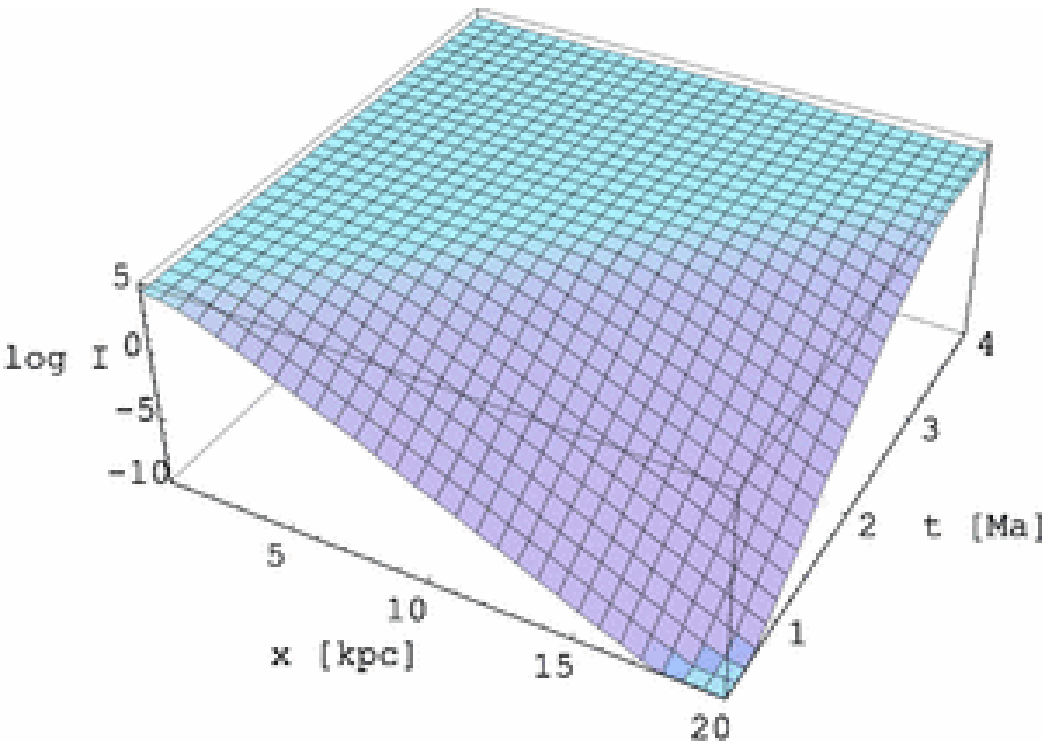}}
\scalebox{0.7}{\includegraphics{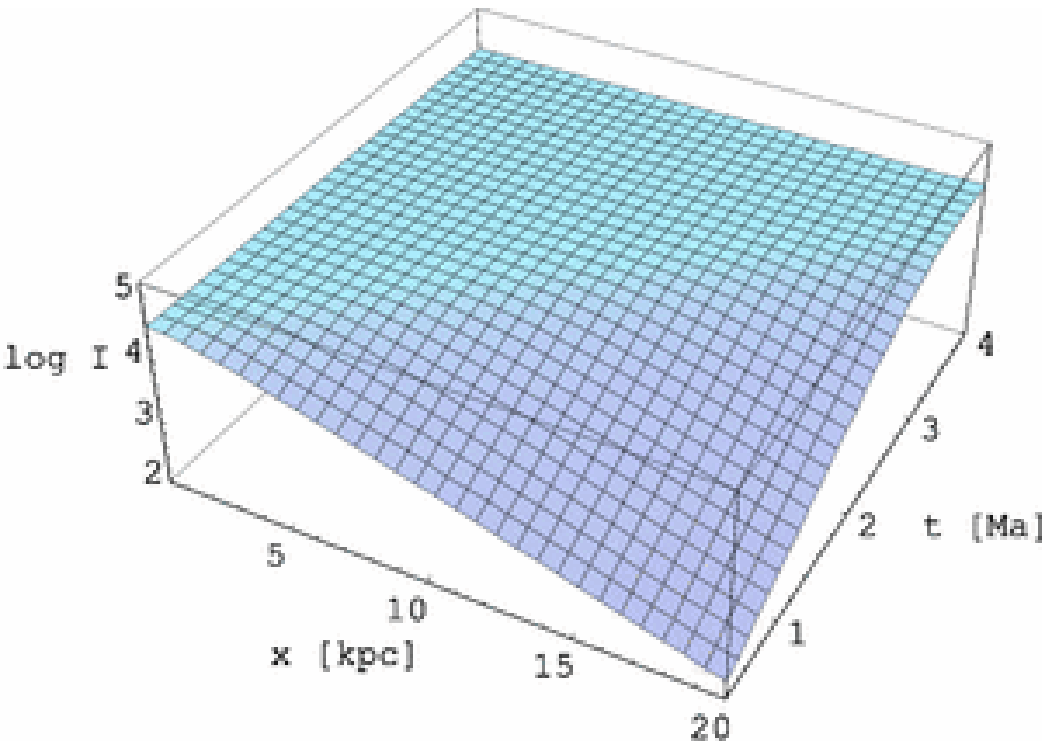}}
\scalebox{0.7}{\includegraphics{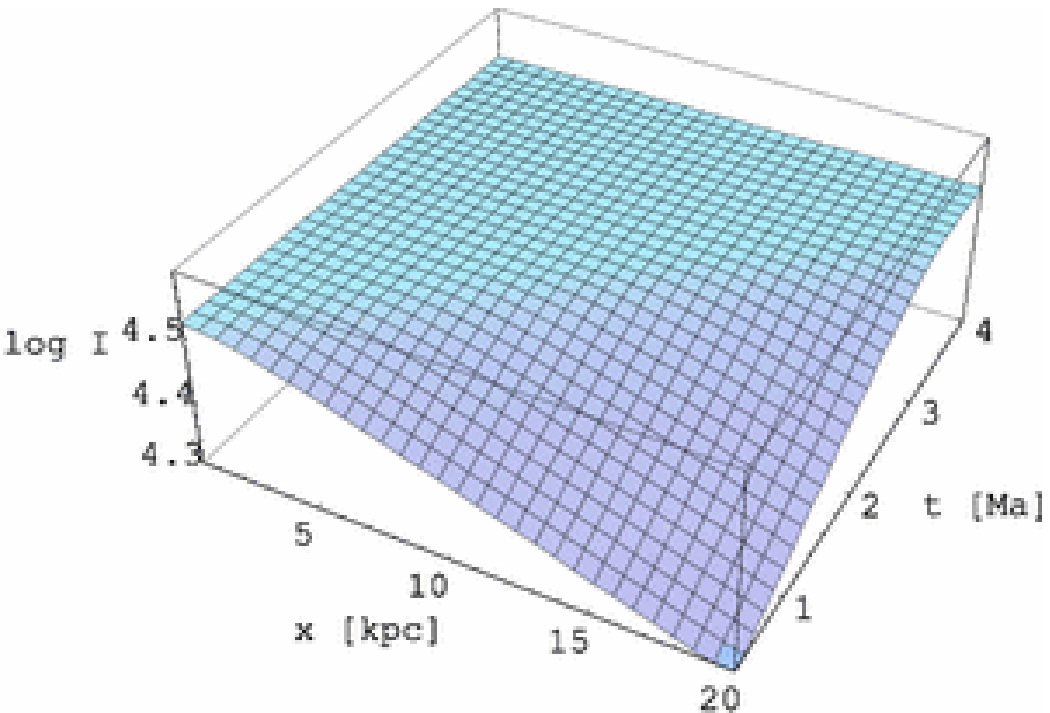}}
\caption{
\label{multifreq}
(a) (top left): The time evolution of the logarithm of the number density of neutral
 hydrogen at three different times
 $t= 0, 2, 4$~Ma. The parameters are as for the standard
  Model b except that three continuum 
levels ($\lambda= 911, 422  , 196$   \AA) are included:
(b, c, d) (top right, bottom left and right): The spatial distribution and time evolution of the intensity of 
radiation
at the the three wavelengths $\lambda= 911, 422 , 196$ \AA) respectively }
\end{figure}

We show in Figure \ref{multifreq} the effect of including three continuum levels 
with
$\lambda = 911, 422$  and $196$ \AA ~in Model b  discussed in Figure \ref{unifden} b. The incident spectrum is 
assumed 
to be flat in wavelength but have the same total average intensity 
 $I_{irr} = 10^5$ erg ~cm$^{-2}$ s$^{-1}$. The spatial profiles of the neutral
density component and its time evolution are markedly different even though the
maximum degree of ionisation remains the same.  We see that the central regions
ionise more quickly and that a well defined front that propagates outwards 
can be seen within our computational region.
The intensity of radiation (Figures \ref{multifreq} b, c and d)  is now strongly
frequency dependent reflecting the $\nu^{-3}$ dependence of the photoionisation
cross-section, with the highest frequency showing the strongest attenuation
close to the source at early times. A distant observer will thus see the source
first at high frequencies, and then at lower frequencies as the Lyman limit
is reached.

\subsection {Sensitivity to input parameters}

The ionisation state of the intergalactic medium depends   
on the nature of the sources of ionising radiation (modelled by $I_{irr}$),
their distribution in space and time, and the density structure of the ambient intergalactic
medium. Our calculations show that small uncertainties in these input parameters 
can have large effects on our conclusions on the state of ionisation of the 
intergalactic medium.

\begin{figure}
\scalebox{0.3}{\includegraphics{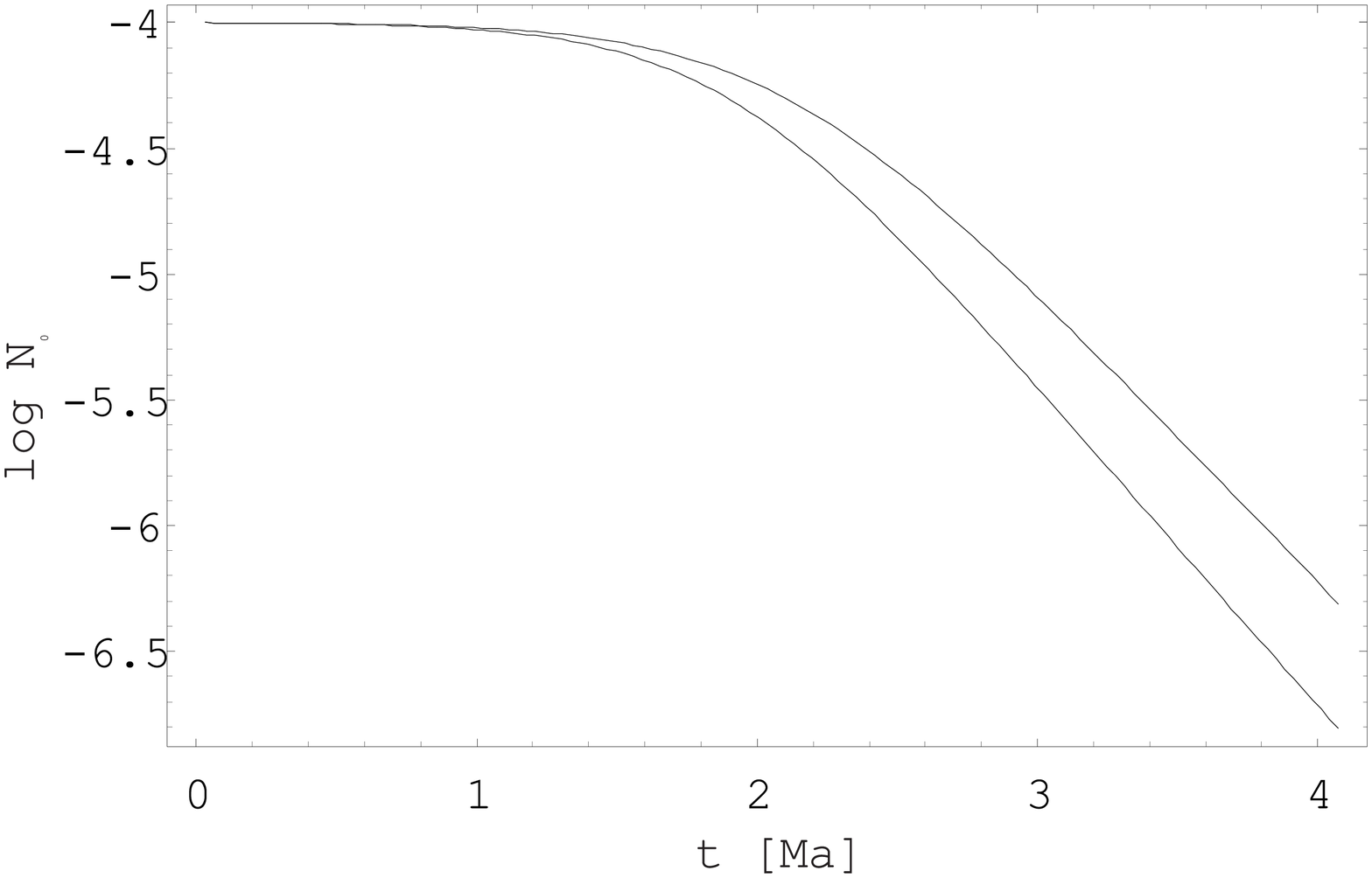}}
\vspace*{0.5cm}
\scalebox{0.3}{\includegraphics{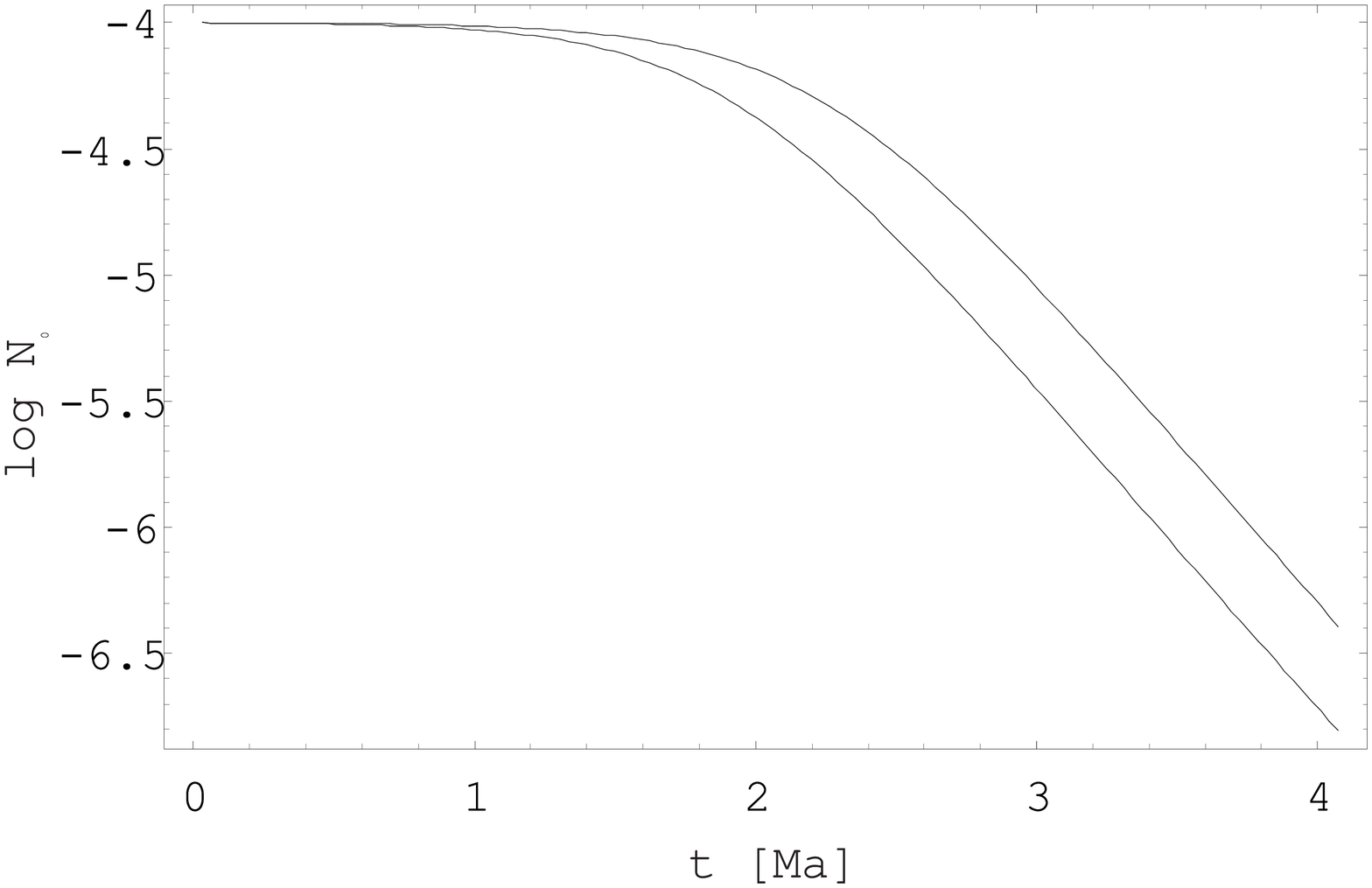}}
\caption{
\label{stochastic}
The sensitivity of the neutral density to a reduction in the ambient density by
$10\%$ (upper panel) and to a reduction  in the incident intensity by $10\%$ (lower panel)
for Model b4.}
\end{figure}

To illustrate the sensitivity of our results on the state of ionisation 
to input parameters, we have repeated the calculations for Model b4, first 
by keeping $I_{irr}$ fixed, and decreasing the ambient density by 10 percent,
and then by keeping the ambient density fixed, and decreasing $I_{irr}$ by 10 percent.
The results of these two sets of calculations are shown in 
Figure \ref{stochastic}. A 10 percent variation in either the density, or
the incident intensity translates to a a factor 3 in the neutral density 
on a time scale of 4 million years. These results show that the details of 
how re-ionisation proceeds requires accurate calculation of the density structure 
in the early universe, such as those that can be obtained from hydrodynamic simulations 
(Abel et al. (1999)).

\section{Model results for multiple sources}

In this section we present  results of one set 
of calculations that were carried out with two sources using the 
generalisation of the numerical method described in section 2.2 (Model s1 in 
Table 1). 
In these models, 
the first source is located at $x=0$,turns on at $t=0$, and  
shines for $5$~Ma after which time it shuts off.
A second source located at $x=20$~kpc turns on at a subsequent time $t=15$~Ma, 
shines for another $5$~Ma, and then shuts off. Each source has
$I_{irr}=10^5$ erg~cm$^{-2}$ s$^{-1}$ and we assume a density distribution with  
a density enhancement centered at $x_s=0$ as in Model b4.

\begin{figure}
\scalebox{0.9}{\includegraphics{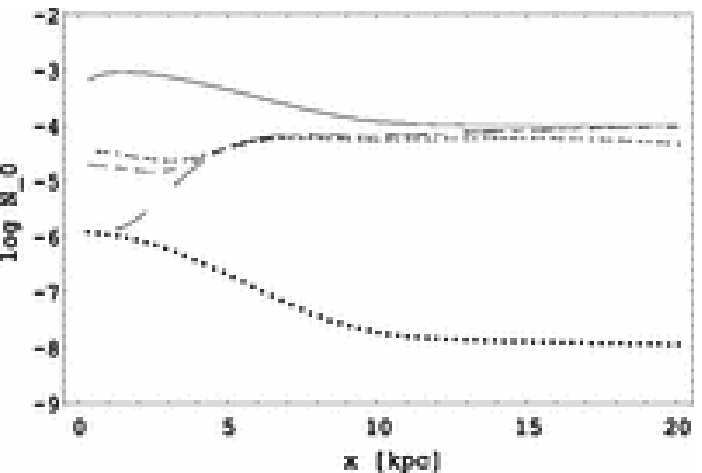}}
\scalebox{0.7}{\includegraphics{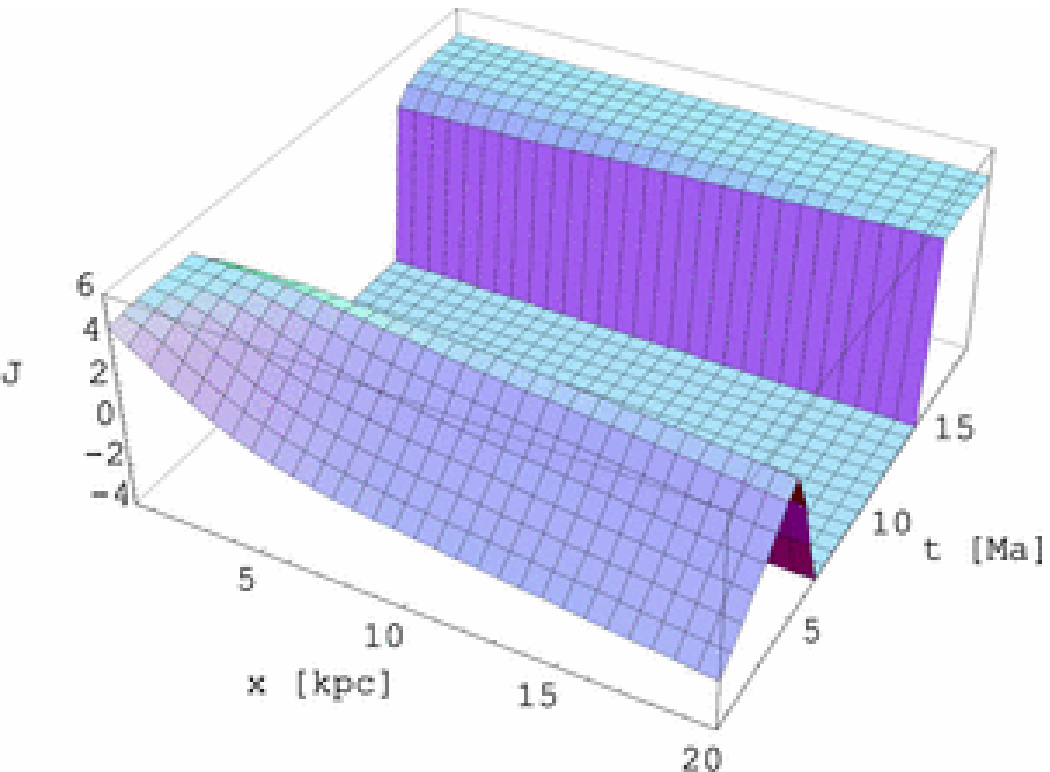}}
\caption{
\label{multisource1_den}
(a) The time evolution of the logarithm of the number density of neutral
hydrogen at five different times $t= 0.6 $~Ma (solid),$t=5 $~Ma (long dashed),
$t=10 $~Ma (short dashed),$t=15 $~Ma (dashed - dotted) and $t=20 $~Ma (dotted)
 for the two source model (Model s1). (b) The mean intensity in (x,t) space for Model s1.}
\end{figure}

 We show in Figure 
\ref {multisource1_den}a the time evolution 
of the logarithm of the number density of neutral hydrogen. Initially, the
behaviour is similar to the single source case, with the ionisation front 
propagating outwards with time starting at $x=0$ with the number density of 
neutral hydrogen decreasing with time at any given spatial position. 
Once the first source of radiation turns off ($t=5$~Ma), the gas begins to
recombine slowly, and the number density of neutral hydrogen increases at any given spatial
location, but not significantly in the intervening $15$~Ma. The second
source then turns on and ionisation proceeds again, but starting from a medium 
which is already partially ionised. In the next $5$~Ma the degree of 
ionsiation drops rapidly reaching 
values as low as $\sim 10^{-5}$ over most of the region at the end of the
calculation. 

The evolution of mean intensity $J$ in our computational volume is shown in 
\ref {multisource1_den}b. If we were located at $x=10$~kpc,
mid way between the two sources, we would first see the radiation from 
the first source rise to reach a peak at $t= 5$~Ma, and the source will
then vanish from view almost instantaneously on a time scale corresponding 
to the light travel time from the source. The second source will then 
come into view on a similar time scale, and brighten slowly as the medium 
becomes significantly ionised again. 


The net effect of having two sources of 
ionisation that turn on at different times is to ionise a significantly
larger volume of gas to a higher degree of ionisation. These calculations
demonstrate that after the first sources turn on, the universe will ionise 
very quickly as subsequent sources turn on, even though the lifetime of 
the sources of radiation may be much shorter than the time between the
sources.

\section {Discussion and Conclusions}
 
We have presented a robust numerical scheme that is unconditionally stable and 
allows the calculation of the time dependent evolution of ionisation domains with 
full time and spatial resolution. The method allows for the explicit inclusion 
of the time dependence of the radiative intensity in the transfer equation.


Our illustrative 1-D examples were chosen to mimic conditions in the early 
universe and highlight the basic processes that are involved in the development 
of ionisation domains around UV sources of radiation. The 
physics of the non-LTE ionisation process is of course well understood, 
and has been incorporated at different levels of approximation in many previous
investigations of cosmological reionisation. As a source first turns on, 
and illuminates a neutral medium, photoionisations will dominate  everywhere 
except that they will 
be less effective in the outer regions because of the larger optical depth 
towards the illuminating 
source. The neutral density will therefore initially decline at different
 rates 
at different spatial 
locations in the medium with the  neutral density in the  inner regions 
declining more rapidly.  
As the neutral density decreases, recombinations become relatively more 
important and slows down 
the rate of decline. In addition the ionisation of the regions closest to the 
source makes this gas more transparent 
enhancing the photoionisation of adjacent outer regions. These two processes 
combine to yield an
expanding ionisation `front'  that  separates the mainly ionised and mainly 
neutral gas. The front
expands at a rate that is determined by the local density and the intensity 
$I_{irr}$ of the source of 
irradiation. As $t$ increases, the amplitude of $\vert \frac{dN_{0}}{dt}\vert$ 
decreases, but an 
equilibrium situation is not reached in any of our calculations within the 
lifetime of the source
($\sim 4 $ Ma).


The above simple picture can change significantly when density fluctuations are 
taken into
consideration. We have investigated the effects that come into play when an 
outward propagating
ionising front
encounters a Gaussian density enhancement of given peak density and half width. 
The density
enhancement slows down the speed of the front in a complex manner depending on 
both of 
these parameters.  Before the front encounters
the enhancement, it has a speed that is nearly constant and equal to the value 
appropriate
to the ambient mean density and the input irradiating intensity $I_{irr}$. As 
 the front
emerges from the density enhancement, the speed  reaches a value 
that is generally lower than before the encounter.  The density enhancement 
`delays' the
propagation of the ionisation front by a time $\Delta t_f$ that is strongly 
dependent
on the peak density and the half width of the assumed Gaussain density 
enhancement.
 Our calculations have shown that delays of
the order of a few tens of million years are possible even for modest 
density enhancements. The delay could therefore exceed the lifetime of the 
ionising
source, so that the density enhancement in effect stops the propagation of the 
front. Due to the long recombination times, such regions 
will remain ionised, enshrouded by neutral gas, until the next source of 
radiation turns on in its vicinity and continues the process of ionsiation.
Our calculations have shown that even though the lifetime of a source of ionising radiation 
may be much smaller than the time between the emergence of such sources, ionisation will proceed 
very effectively once started.

Various generalisations, such as extension to three dimensions are easily
implemented. Helium ionisation can be readily incorporated by the use of more levels (and therefore more frequency
points) and the cosmological expansion term by implicitly discretizing 
the frequency variable. Our future plans are to apply our numerical method to 
outputs of cosmological simulations.

\section*{Acknowledgements}
The authors thank the Deutsche Forschungsgemeinschaft (SFB 439, project A4) for 
financial support and Owen Dive for reading through the manuscript and providing 
valuable comments.

\end{document}